\documentclass[10pt,flushrt,preprint]{aastex62}

\usepackage{graphicx}
\usepackage[space]{grffile}
\usepackage{latexsym}
\usepackage{textcomp}
\usepackage{longtable}
\usepackage{multirow,booktabs}
\usepackage{amsfonts,amsmath,amssymb}
\usepackage{natbib}
\usepackage{url}
\usepackage{hyperref}
\hypersetup{colorlinks=false,pdfborder={0 0 0}}
\newif\iflatexml\latexmlfalse
\DeclareGraphicsExtensions{.pdf,.PDF,.png,.PNG,.jpg,.JPG,.jpeg,.JPEG}

\usepackage[utf8]{inputenc}
\usepackage[english]{babel}
\usepackage{graphicx}

\newcommand\tractor{\emph{The\,Tractor}}%

\submitjournal{Astronomical Journal (AJ)}

\shorttitle{Overview of the DESI Legacy Imaging Surveys}
\shortauthors{DESI Imaging Team}

\newcommand{\noao}{National Optical Astronomy Observatory, 950 N. Cherry Ave.,
Tucson, AZ 85719}
\newcommand{\lbnl}{Lawrence Berkeley National Laboratory, 1 Cyclotron Rd., Berkeley, CA 94720}
\newcommand{\berkeley}{Department of Physics, University of California at Berkeley, Berkeley, CA 94720}
\newcommand{\steward}{Steward Observatory, University of Arizona, 933 N. Cherry Ave., Tucson, AZ 85721}
\newcommand{\harvardcfa}{Harvard-Smithsonian Center for Astrophysics,
  60 Garden St., Cambridge, MA 02138}
\newcommand{\ctio}{Cerro Tololo Inter-American Observatory, National Optical Astronomy Observatory, Casilla 603, La Serena, Chile}
\newcommand{\wyoming}{Department of Physics \& Astronomy, University  of Wyoming, 1000 E. University, Dept 3905, Laramie, WY 8207}
\newcommand{\fermilab}{Fermi National Accelerator Laboratory, P.O. Box 500, Batavia, IL 60510}
\newcommand{\icecsic}{Institute of Space Sciences (ICE, CSIC), Campus UAB, Carrer de Can Magrans, s/n,  08193 Barcelona, Spain}
\newcommand{\ieec}{Institut d'Estudis Espacials de Catalunya (IEEC), 08193 Barcelona, Spain}
\newcommand{\toronto}{Department of Astronomy \& Astrophysics, University of Toronto, Toronto, ON M5S 3H4, Canada}
\newcommand{\siena}{Department of Physics and Astronomy, Siena College, 515 Loudon Rd., Loudonville, NY 12211}
\newcommand{\irfucea}{IRFU, CEA, Universit\'e Paris-Saclay, F-91191 Gif-sur-Yvette, France}
\newcommand{\naoc}{Key Laboratory of Optical Astronomy, National Astronomical Observatories, Chinese Academy of Sciences, Beijing 100012, China}
\newcommand{\ucirvine}{Department of Physics and Astronomy, University of California, Irvine, Irvine, CA 92697}
\newcommand{\rochester}{Department of Physics and Astronomy, University of Rochester, 500 Wilson Blvd., Rochester, NY 14627}
\newcommand{\durham}{Centre for Extragalactic Astronomy, Durham University, South Rd., Durham, DH1 3LE, UK}
\newcommand{\stsci}{Space Telescope Science Institute, 3700 San Martin Dr., Baltimore, MD 21218}
\newcommand{\warwick}{Department of Physics, University of Warwick, Coventry CV4 7AL, UK}
\newcommand{\epfl}{Institute of Physics, Laboratory of Astrophysics, Ecole Polytechnique F\'ed\'erale de Lausanne (EPFL), Observatoire de Sauverny, 1290 Versoix, Switzerland}
\newcommand{\smu}{Department of Physics, Southern Methodist University, 3215 Daniel Ave., Dallas, TX, 75205}
\newcommand{\ohio}{Department of Physics and Astronomy, Ohio University, Clippinger Labs, Athens, OH 45701}
\newcommand{\ucl}{Department of Physics and Astronomy, University College London, London WC1E 6BT, UK}
\newcommand{\ccapp}{Department of Astronomy and Center for Cosmology and Astroparticle Physics, The Ohio State University, Columbus, OH 43210}
\newcommand{\unam}{Instituto de F\'isica, Universidad Nacional Aut\'onoma de M\'exico, A.P. 20-364, 04510 Ciudad de M\'exico, M\'exico}
\newcommand{\sorbonne}{Sorbonne Universit\'e, Universit\'e Paris-Diderot, CNRS-IN2P3, Laboratoire de Physique Nucl\'eaire et de Hautes Energies, LPNHE, F-75005 Paris, France}
\newcommand{\kansas}{Department of Physics and Astronomy, University of Kansas, 1251 Wescoe Hall Dr., Room 1082, Lawrence, KS 66045}
\newcommand{\pitt}{Department of Physics and Astronomy and PITT PACC, University of Pittsburgh, 3941 O’Hara St., Pittsburgh, PA 15260}

\begin{document}

\title{Overview of the DESI Legacy Imaging Surveys}

\author{Arjun Dey}
\affiliation{\noao}
\email{dey@noao.edu}

\author{David J. Schlegel}
\affiliation{\lbnl}

\author{Dustin Lang}
\affiliation{Dunlap Institute, University of Toronto, Toronto, ON M5S 3H4, Canada}
\affiliation{\toronto}
\affiliation{Perimeter Institute for Theoretical Physics, Waterloo, ON N2L 2Y5, Canada}

\author{Robert Blum}
\affiliation{\noao}

\author{Kaylan Burleigh}
\affiliation{\lbnl}

\author{Xiaohui Fan}
\affiliation{\steward}

\author{Joseph R. Findlay}
\affiliation{\wyoming}

\author{Doug Finkbeiner}
\affiliation{\harvardcfa}

\author{David Herrera}
\affiliation{\noao}

\author{St\'{e}phanie Juneau}
\affiliation{\noao}

% ...

\begin{abstract}
The \href{http://legacysurvey.org/}{DESI Legacy Imaging Surveys} are a combination of three public projects (the Dark Energy Camera Legacy Survey, the Beijing-Arizona Sky Survey, and the Mayall $z$-band Legacy Survey) that will jointly image $\approx$14,000~deg$^2$ of the extragalactic sky visible from the northern hemisphere in three optical bands ($g$, $r$, and $z$) using telescopes at the Kitt Peak National Observatory and the Cerro Tololo Inter-American Observatory. The combined survey footprint is split into two contiguous areas by the Galactic plane. 
The optical imaging is conducted using a unique strategy of dynamically adjusting the exposure times and pointing selection during observing that results in a survey of nearly uniform depth.
In addition to calibrated images, the project is delivering a catalog, constructed by using a probabilistic inference-based approach to estimate source shapes and brightnesses. The catalog includes photometry from the $grz$ optical bands and from four mid-infrared bands (at 3.4$\mu$m, 4.6$\mu$m, 12$\mu$m and 22$\mu$m) observed by the {\it Wide-field Infrared Survey Explorer} ({\it WISE}) satellite during its full operational lifetime. The project plans two public data releases each year. All the software used to generate the catalogs is also released with the data. This paper provides an overview of the Legacy Surveys project. 

\end{abstract}
\keywords{surveys -- catalogs}

\section{Introduction}

Explorations of the universe begin with images. In the last few decades, systematic surveys of the sky across the electromagnetic spectrum have revolutionized the ways in which we study  physical processes in known astronomical sources, identify new astrophysical sources and phenomena, and map our environs \citep[e.g., see][ for an excellent summary]{skysurveys}. The amazing bounty of wide-field imaging surveys at optical wavelengths has been recently demonstrated by the Sloan Digital Sky Survey \citep[SDSS;][]{,york2000,sdssDR7,sdssDR8},  Pan-STARRS1 \citep[PS1;][]{panstarrs} and the Dark Energy Survey \citep{des}, all of which continue to advance our knowledge of the universe in multiple fields of astrophysics \citep[e.g.,][]{desplus}.

In this paper we describe the DESI Legacy Imaging Surveys (hereafter The Legacy Surveys) aimed at mapping 14,000~deg$^2$ of the extragalactic sky in three optical bands ($g$, $r$ and $z$). The very wide areal coverage and the need to finish the survey in less than three years necessitated the use of three different telescope platforms: the Blanco telescope at the Cerro Tololo Inter-American Observatory;  the Mayall Telescope at the Kitt Peak National Observatory; and the University of Arizona Steward Observatory 2.3m (90inch) Bart Bok Telescope at Kitt Peak National Observatory. 
In addition, the Legacy Surveys source catalogs incorporate mid-infrared photometry for all optically-detected sources from new image stacks of data from the {\it Wide-field Infrared Survey Explorer} satellite \citep{WISE}. 

\section{Motivation for a New Wide-Field Imaging Survey \label{sec:motivation}}

\subsection{Imaging for the Dark Energy Spectroscopic Instrument Surveys}

The Legacy Surveys are motivated by the need to provide targets for the Dark Energy Spectroscopic Instrument (DESI) survey.  DESI is an international project that is constructing a
5000-fiber multi-object spectrograph for the Mayall 4m telescope at the Kitt Peak National Observatory \citep{desiInstrument}.
Over a five-year period (2019--2024), DESI will measure the redshifts of 35 million galaxies and quasars,  including $\sim700,000$ QSOs at $z>2.1$\footnote{We shall use the terms ``quasar'' and ``QSO'' interchangeably throughout this paper.} suitable for probing the structure of the intergalactic medium at high redshift  \citep{desiScience}.  The DESI Key Project will use these maps of the large scale matter distribution traced by galaxies and the Lyman-$\alpha$ forest to measure the expansion history of the universe over the past 10 billion years. The goal is to provide sub-percent accuracy
constraints on the equation of state of dark energy and its time evolution \citep[cf.][]{bosscosmology2017}. The DESI project will also provide precise constraints on the growth of structure in the universe by using measurements of redshift-space distortions \citep[e.g.,][]{guzzo2008,blake2011,pezzotta2017}. In order to reach percent-level precision on the cosmological parameters, the DESI survey requires spatially dense samples of galaxy and QSO tracers across very large areas of the sky ($>$10,000~deg$^2$).  The SDSS and PS1 surveys are both too shallow to reliably select the DESI targets, and the contiguous extragalactic (i.e., at $\vert b\vert\ge 15^\circ$) SDSS footprint is too small. The DES survey reaches adequate depth, but covers only 5000~deg$^2$ mainly in regions too far south to be reached from Kitt Peak. These considerations motivated the Legacy Surveys, which are deeper than SDSS and PS1 and cover a much larger area than DES in the northern sky. Imaging for the Legacy Surveys is on track to be completed prior to the start of the DESI spectroscopic survey in 2019.  The detailed requirements placed by the DESI target selection on the imaging surveys are described in more detail in an Appendix to this paper (see \S~A1). 

\subsection{Complementing Existing Spectroscopy}

Beyond the primary goal of providing DESI targets, the imaging survey described in this paper has more wide-ranging astrophysical motivations.  The Sloan Digital Sky Survey \citep[SDSS; e.g.,][]{sdssDR7,sdssDR14} project has overwhelmingly demonstrated the power of combining wide-field imaging and spectroscopic surveys within the same footprint. The SDSS-I,II,III/BOSS surveys contain $\sim$2.8 million spectra, including 300,000 unique stars, 700,000 galaxies at $z<0.2$, 500,000 galaxies at $0.2 < z < 0.5$, 1 million galaxies at $z>0.5$, 100,000 QSOs at $z<2$, and 200,000 QSOs at $z>2$.\footnote{{\tt http://sdss3.org}} The median extragalactic redshift of these samples is already $z_{\rm med}\approx0.5$, and SDSS-IV/eBOSS \citep[][2014--2020]{Daw16} is currently adding another 600,000 galaxies at $0.6<z<1$ \citep{Pra16,raichoor2017} and 500,000 new QSOs at $z>0.9$ \citep{Mye15}.  Most of these data are already available publicly. 

While most SDSS-I spectra targeted nearby galaxies
($r<17.77$; i.e., 4--5 magnitudes brighter than the imaging detection limit),
BOSS (SDSS-III) targeted much fainter sources (galaxies to
$i=19.9$ and QSOs to $g=22$), near the limits of the
original SDSS imaging (Dawson et al.~2013);  eBOSS (SDSS-IV) goes even fainter \citep[see, e.g.,][]{sdssDR14}..
While adequate for the study of large-scale structure, the full
science impact of these data is limited by the depth and quality
of the existing imaging. The bulk of existing spectroscopic redshifts are in the northern sky and have poor overlap with most deep, wide-field imaging surveys (see Figure~\ref{fig:zsfootprint}).  The SDSS imaging data (which provided the spectroscopic targets) do not provide precise photometry, well-resolved size measurements, detailed morphologies, or
environmental measures for the bulk of the faint galaxies targeted by the existing spectroscopy. 

The Legacy Surveys will greatly remedy this situation by imaging the entire BOSS footprint to magnitudes suitable for the study of the
$z>0.5$ universe (see Figure~\ref{fig:zsfootprint}).  
 Based on the magnitude distribution of galaxies in the $z$COSMOS catalog \citep{zCOSMOS}, imaging to the 5$\sigma$ $z$-band depth of the Legacy Surveys will result in increasing the number of detected $z>0.5$ galaxies ($z>1$) galaxies by a factor of $>$15 ($>200$) over SDSS.
Measuring $g-r$ vs. $r-z$ colors cleanly
isolates $z>0.5$ galaxies. Optical photometry coupled with the {\it WISE} mid-infrared photometry
can be used to measure stellar masses and AGN activity for such galaxies \citep[see, e.g., \S~3 of ][and references therein]{desiScience}. 
can be used to resolve morphologies and structural parameters for all SDSS spectroscopic galaxies.
The combination of the image quality (median FWHM in the $z$ band of $\approx$1.1\arcsec) and depth of the Legacy Surveys can be used to measure improved morphologies and structural parameters for all SDSS spectroscopic galaxies.

Spectroscopy complements deep imaging; it provides: robust redshifts; a crisp 3-d view of
large-scale structure; dynamical information through velocity
dispersions; spectral diagnostics of stellar populations, star
formation rates, and nuclear activity; and probes 
of the intergalactic medium through absorption line studies. The combination enables numerous astrophysical studies. For example: 

\begin{itemize}
\item{{\it The Evolution of Galaxy Clusters:}
While SDSS has obtained redshifts of 1.5 million massive galaxies, often
the central, brightest galaxies in groups and
clusters, current imaging often cannot detect their satellites.
The Legacy Surveys will significantly improve stellar mass
models for these galaxies and enable a sensitive search for faint cluster
members. Extrapolating from the SDSS
Stripe 82 imaging \citep{rykoff2014,rykoff2016}, we expect to identify $\sim$75,000 clusters, nearly
all of which will have spectroscopic redshifts available from SDSS.
Spectroscopy provides three key benefits not available to photometric-only
surveys: 1) calibration of cluster masses by stacked velocity dispersion
measurements \citep[e.g.,][]{becker2007}; 2) tests of general relativity by the comparison of
the velocity field around clusters to the weak lensing shear mass profile
\citep[e.g.,][]{lam2012,zu2014};
and 3) calibration of cluster masses by detecting the weak-lensing magnification of the luminosity function of background galaxies and quasars \citep{coupon2013,coupon2015}. Magnification-based methods
have systematic uncertainties that are completely independent
from the shape and photometric redshift systematics expected to dominate the
error budget of imaging-only surveys like DES or LSST, thereby
enabling a critical consistency test with these surveys.}

\item{{\it Galaxy Halos Through Cosmic Time:}
The contents (and shapes) of galaxy dark matter halos can be revealed
from the cross-correlation of spectroscopic and imaging maps \citep{eisenstein2005,tal2013}
and from galaxy-galaxy weak lensing \citep[e.g.,][]{mandelbaum2016}.
These methodologies benefit substantially from deeper
imaging, 
with statistical errors on cross-correlations and lensing signals often scaling as $N_{\rm gal}^{-1/2}$.
Higher precision is crucial: variations in clustering as a
function of galaxy properties are often only of order 10\%, so distinguishing
between models requires percent-level clustering measurements.
 The $z\approx22.8$ AB mag 5$\sigma$ depth of the Legacy Surveys imaging will increase the samples available to these methodologies by factors of $>$15 \citep[based on comparisons to the $z$COSMOS catalogs;][]{zCOSMOS}.
Cross-correlation studies use angular correlations to tie deep photometric
catalogs to overlapping spectroscopic maps, measuring the mean environments
and clustering of galaxies and AGN with great accuracy.
SDSS has provided
high-precision results at lower redshift using these techniques, e.g.,
measuring the mean environment of galaxies as a function of luminosity,
color, and scale \citep{hogg2003,eisenstein2005,masjedi2006,jiang2012}
and interpreting this
to constrain halo populations and merger rates \citep{zheng2009,watson2012}.
The Legacy Surveys will extend this to far
larger ($>$10--100$\times$) spectroscopic and photometric samples
at high redshift, measuring the satellite distributions around 
central galaxies as a function of redshift, luminosity, stellar
mass, color, major axis orientation, velocity dispersion, [\ion{O}{2}]
emission line equivalent width, etc.  Cross-correlation also enables more robust
clustering measurements around rare spectroscopic populations, and the ability to
calibrate galaxy redshift distributions from imaging data \citep{newman2008,Mye09,menard2013,schmidt2013}.}

\item{{\it The Evolution of Halo Gas:} SDSS spectra have already yielded $>$50,000 MgII absorption line systems at $0.4<z<2.5$ toward background QSOs \citep{zhu2013}, and eBOSS will
increase the number of sightlines to nearly a million.
By cross-correlating 2,000 absorbers at $z\sim 0.5$ with SDSS
photometric galaxies, \citet{lan2014} extracted new relations between galaxy
properties and their surrounding gas (e.g., their Fig.\ 2a).
The Legacy Surveys will dramatically improve this type of analysis by
extending its reach from $z\sim 0.5$ to $z\sim 2$,
sampling the full range of $\sim$100,000 identified absorbers.
This will map the cosmic evolution of halo gas as a function of redshift,
making it possible to understand its dependence on
galaxy type, orientation, luminosity, star-formation rate, environment, etc.}

\item{{\it The Halo of the Milky Way:} The SDSS, PS1 and DES imaging surveys have revolutionized the study of the Milky Way,
finding numerous stellar halo streams \citep[e.g.,][]{newberg2002,yanny2003,grillmair2009,bernard2016,shipp2018} and dwarf
galaxies \citep{willman2005,laevens2014,dw2015,bechtol2015}.
The Legacy Surveys will map at least twice as far out
into the Galactic halo over 14,000~deg$^2$, increasing
the volume of the MW explored by a factor of ${\sim}5$ relative
to SDSS+Pan-STARRS. This will enable tests of predictions that stellar halo substructure dramatically increases with distance \citep{bell2008,helmi2011}.
Our photometric parallax-based maps will extend to $\sim 40$~kpc using
main sequence stars \citep{ivezic2008,juric2008},
$\sim 80$~kpc using $gr$-selected main sequence turnoff stars
\citep{bell2008}, and $\sim 150$~kpc using $ugr$-identified Blue Horizontal Branch (BHB) stars
where $u$-band is available \citep{ruhland2011}.
The deeper data on known streams \citep{odenkirchen,carballo2018} will be used to test for the presence
of ``missing satellites'' via their signatures
in these streams \citep{carlberg2009,yoon2011}.
Imaging from the Legacy Surveys should be sufficient to discover $8$--$20$ new dwarf galaxies.
Each dwarf galaxy discovery immediately adds years of Fermi integration
to the search for dark matter detection via gamma rays \citep{fermilat2017}.
Finally, given the 10-year time baseline between imaging from 
SDSS and the Legacy Surveys, proper motions should be measured to accuracies of a few
milliarcsec per year for stars 2~mag fainter than the Gaia limits.}

\end{itemize}

\subsection{Photometry from the {\it WISE} Satellite}
The Legacy Surveys will greatly enhance in the utility of
the mid-IR imaging data from the {\it WISE} satellite by providing deep
template $grz$ optical images for matched photometry using \tractor\
package \citep[see \S~\ref{sec:tractor}]{lang16code}.  By optimally matching
{\it WISE} to deep optical imaging, one can partially deblend the images of confused {\it WISE} sources
and improve the signal-to-noise ratio of their mid-infrared photometry and color measurements. 
Using SDSS $r$-band templates already shows substantial improvement,
but the deeper Legacy Surveys images will allow extraction of fainter, higher redshift sources.
The extended {\it WISE} mission will more than quadruple the exposure time of the original {\it WISE} all-sky survey (cf. the AllWISE catalog) in the
3.4 and 4.6 $\mu$m bands by the end of 2018 and provide multiple epochs for identification of mid-infrared variable sources.  The Legacy
Surveys will provide matched {\it WISE} mid-infrared photometry for hundreds of millions of optical
sources.  Properly matched optical-to-mid-IR photometry will allow
more robust estimation of stellar masses and improved photometric
redshifts for extragalactic objects.  Such photometry will also facilitate high-fidelity selection of massive galaxies
to $z\sim1.5$--$2$, and the selection of nearly all optically detected quasars.

\section{Footprint}
\label{sec:footprint}

The footprint of the Legacy Surveys is designed to correspond to the DESI Survey footprint, which is defined to be the extragalactic sky above a Galactic latitude of 
$b=15^\circ$ that can be observed spectroscopically from Kitt Peak (i.e., at declination $\delta > -20$$^{\circ}$). These selections result in an 
$\approx$14,000 deg$^2$ area, which contains two 
contiguous regions, one in the North Galactic Cap (NGC)
covering 9900 deg$^2$ 
and one in the South Galactic
Cap (SGC) covering 4400 deg$^2$ 

The basic criteria described result in a larger area in the NGC (A semester) relative to the SGC (B semester). However, we also need to conduct a uniform, wide-area extragalactic survey with fields that can be scheduled throughout the year, minimizing observations at high airmass (at low or high declinations) and
in regions of high Galactic extinction or high stellar density. To minimize scheduling issues for DESI, the NGC portion of the footprint is trimmed to declination $\delta > -8.2^\circ$ and the SGC area extends southward to $\delta\approx-13.3^\circ$ in regions not covered by the Dark Energy Survey \citep[DES;][]{des}, and to $\delta\approx-18.4^\circ$ in the region covered by DES. These choices were informed by realistic simulations of the DESI survey including a dynamic observing model similar to that described in \S\,\ref{sec:dynamic}.


Since the primary motivation is an extragalactic cosmological survey, additional cuts are imposed to remove those parts of the sky with the largest stellar density.
For the survey regions closest to the Galactic center (i.e., $-90^\circ\,< l < +90^\circ$), only regions with Galactic latitude $|b| > 18^\circ$ are selected; in 
the Galactic anti-center, a less stringent criterion of $|b| > 14^\circ$ is imposed, allowing the survey to extend a bit closer to the Galactic plane.

Finally, the selected footprint is modified to both avoid
small holes within the survey and to avoid largely disconnected
regions that arise as a result of the $E(B-V)$ cuts.
For example, an ``orphaned'' area of 600~deg$^2$ in the northern part
of the SGC has therefore been excluded from the DESI footprint.

\begin{figure}
\includegraphics[scale=0.65]{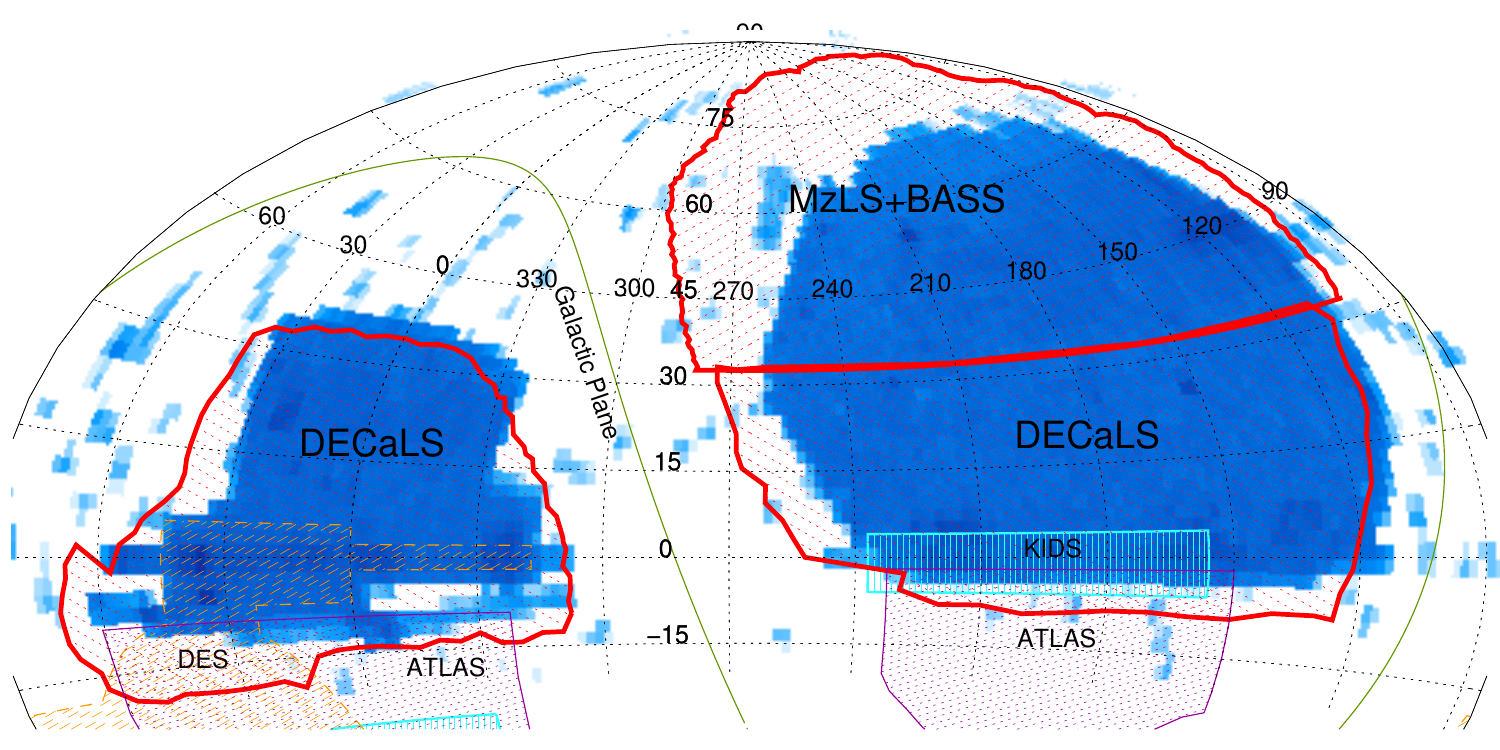}
\caption{The footprints of the optical imaging surveys contributing to DESI imaging,  demarcated by the thick red outlines, 
are shown here in an equal-area Aitoff projection in equatorial coordinates. The region covered by the BASS and MzLS surveys is almost entirely in the North Galactic Cap (NGC) at declinations $\delta\ge+32^\circ$, and DECaLS covers the entire South Galactic Cap and the $\delta\le+34$ regions in the NGC. The regions covered by existing wide-area spectroscopic redshift surveys \citep[SDSS, 2dF, and BOSS;][]{sdssDR7,2dF,sdssDR14} are shown in the blue greyscale in the map above, where the darker colors represent a higher density of spectroscopic redshifts. The Legacy Surveys provide deeper imaging and can leverage the existing spectroscopy in these regions, unlike most other existing or ongoing deep imaging surveys \citep[e.g., DES, ATLAS, KIDS, etc.;][]{des,ATLAS,KIDS}. \label{fig:zsfootprint}}
\end{figure}

The final footprint is shown in Figure~\ref{fig:zsfootprint}. 
The DESI spectroscopic survey is expected to observe most or all this footprint, dependent upon the level of completion of
the Legacy Surveys.

\begin{figure}
\vspace{-1.3in}
\includegraphics[scale=0.65]{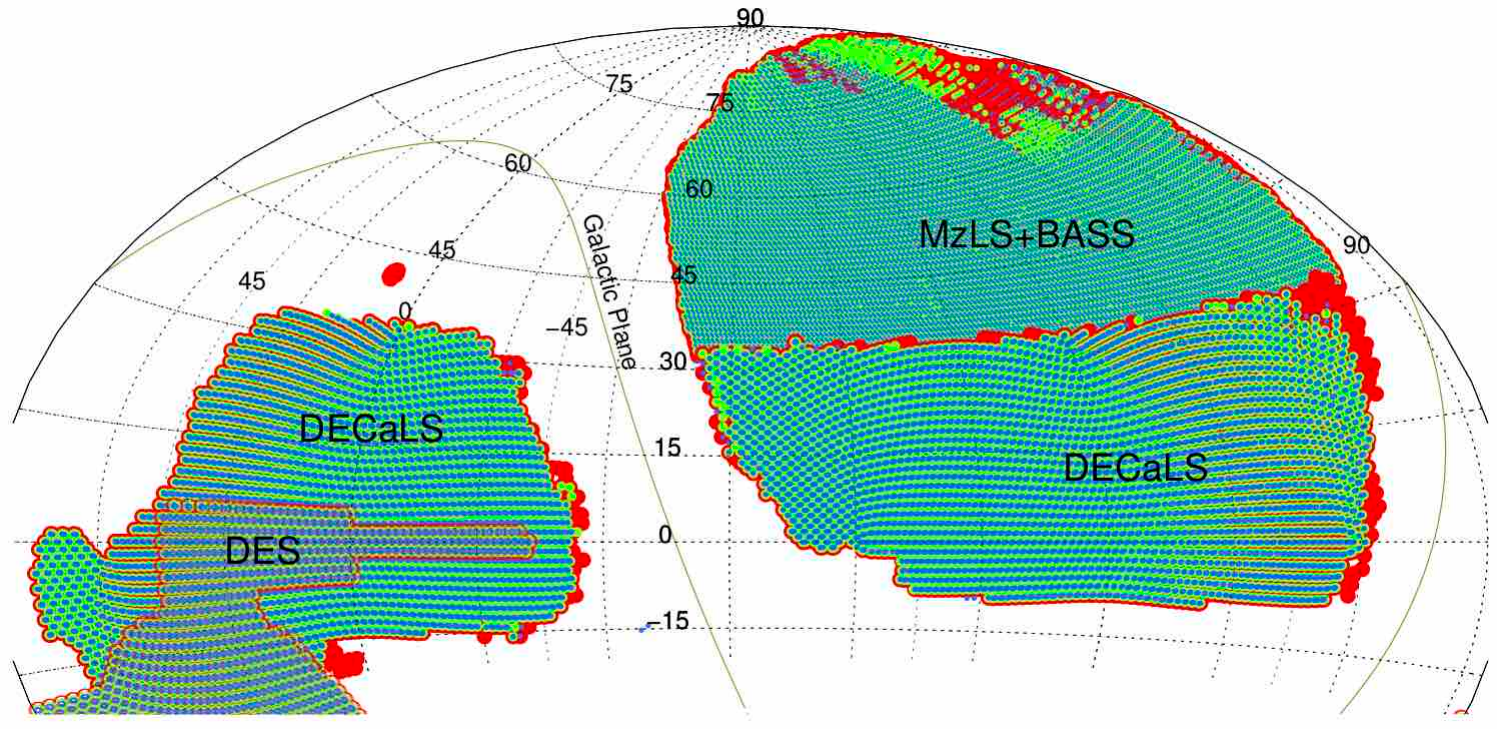}
\vspace{-1.3in}
\caption{The current imaging coverage (as of December 2018) of the Legacy Surveys. Red, green and blue dots represent regions where there is at least a single $z$, $r$ or $g$ band observation, respectively. The MzLS $z$-band survey is now complete; BASS $g$ and $r$-band observations and all DECaLS $grz$ observations will be completed by March 2019}. For a more up-to-date status, see \url{http://legacysurvey.org/status/}.
\end{figure}

\section{The Three Surveys \label{sec:surveys}}

The four target classes that will be used as cosmological tracers by DESI can be selected using a combination of optical imaging data in the $g$, $r$, and $z$ bands and mid-infrared imaging in the 3.4\,$\mu$m and 4.6\,$\mu$m {\it WISE} bands (see \S\,\ref{sec:desirequirements} for further details). DESI requires that the Legacy Surveys deliver 5$\sigma$ detections of a ``fiducial'' $g$=24.0, $r$=23.4 and $z$=22.5~AB mag galaxy with an exponential light profile of half-light radius $r_{\rm half}=0.45$\,arcsec. DESI also requires the depth (and the resulting target selection) to be as uniform as possible across the survey footprint. Ideally, a cosmological survey would use the same imaging data to select all science and calibration targets. However, the ambitious footprint coupled with the short timeline for DESI and lack of very-wide-field imaging capabilities in the northern hemisphere necessitated using multiple platforms to cover the footprint.

Consequently, a combination of three telescopes is used to provide the optical imaging for the Legacy Surveys:
the Blanco 4-m telescope at Cerro Tololo, the Bok 90-inch and the
Mayall 4-m telescope at Kitt Peak (see Table~\ref{tab:surveys}). The areas of the Legacy Surveys
imaged using each of these telescopes are shown
in Figure~\ref{fig:zsfootprint} and the next three subsections discuss
these surveys and their current status in more detail. The status of
the {\it WISE} data used in the Legacy Surveys catalogs is presented in \S\,\ref{sec:WISE}.

DESI targeting requires uniformity in the imaging within each sub-footprint, and resorting to multiple platforms poses challenges. In order to minimize  non-uniformity and cross-calibration issues, the overall footprint was divided into only three contiguous regions. Two of these three regions are being imaged using the Dark Energy Camera on the Blanco telescope, the instrument and telescope combination delivering the widest field of view (and therefore the fastest survey capability). The other region, which is in the NGC north of $\delta\approx+34^\circ$, is being imaged from Kitt Peak using the 90Prime Camera on the Bok telescope for the $g$ and $r$ bands, and the Mosaic-3 camera on the Mayall telescope for the $z$ band observations. The sub-footprints of these individual surveys overlap in the NGC (in an area of $\approx$300~deg$^2$) in the declination range $+32^\circ<\delta<+34^\circ$, so that the color transformations between the different camera+telescope combinations can be calibrated to high precision and accuracy. An additional $\approx$100~deg$^2$ in SDSS Stripe~82 is also being imaged by all three surveys to aid the cross-calibration (see Table~\ref{tab:overlap}). 

A fill factor of unity is not required for the DESI Key Project.
As long as the detailed sky mask is well-characterized, the clustering analyses
can make use of that mask with information loss proportional to this
fractional loss of area.  The DESI requirements are that the coverage
to full depth in all three optical bands should exceed 90\% of the footprint, and that 95\% (98\%) 
must be within 0.3 (0.6) magnitudes of full-depth. The observing nights allocated to each survey are shown in Table~\ref{tab:observing}. 

\begin{figure}[!th]
\centering
\includegraphics[scale=0.65]{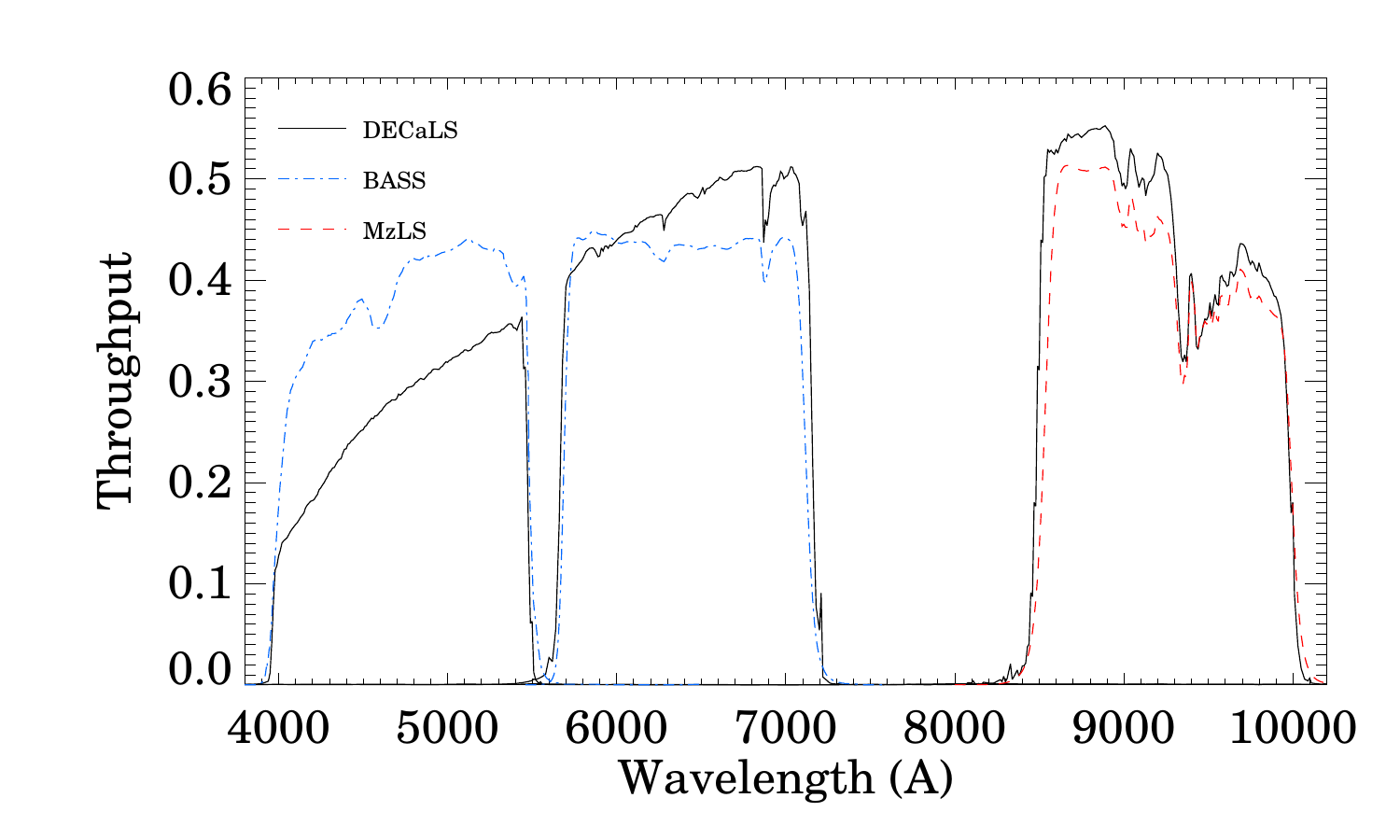}
\caption{The effective band-passes used for the Legacy Surveys. The DECaLS, BASS and MzLS effective filter throughputs for the entire system are shown as solid (black), dashed (blue) and dot-dashed (red) curves, respectively. These include the transmission of the atmosphere (at a median airmass of 1.1 for BASS and MzLS and of 1.4 for DECaLS), the reflectivity and obscuration of the primary mirror, the corrector transmission, and the quantum efficiency of the CCDs. The transmission data are archived on the Legacy Surveys' website at \url{http://legacysurvey.org/dr6/description/} (BASS $gr$ and MzLS $z$) and 
\url{http://legacysurvey.org/dr7/description/} (DECaLS $grz$).
\label{fig:filters}}
\end{figure}

\begin{table}[!th]
\centering
\caption{Telescopes used for the Legacy Surveys}
\begin{tabular}{llccc}
\hline
Survey & Telescope/ &  Bands  & Area & Location \\
 & Instrument & & deg$^2$ &  \\
\hline
DECaLS & Blanco/DECam &  $g$,$r$,$z$ & 9,000 & NGC(Dec~$\le+32\deg$)+SGC \\
BASS & Bok/90Prime& $g$,$r$ & 5,000 & NGC (Dec~$\ge+32\deg$) \\
MzLS & Mayall/Mosaic-3& $z$ & 5,000 &NGC (Dec~$\ge+32\deg$) \\
{\it WISE} \& {\it NEOWISE} & {\it WISE} W1,W2 & 3.4,4.6 $\mu$m & all-sky &all-sky \\
{\it WISE} & {\it WISE} W3,W4 & 12,22 $\mu$m & all-sky &all-sky \\
\hline
\end{tabular}
\label{tab:surveys}
\end{table}

\begin{table}[!th]
\centering
\caption{Regions where Surveys Overlap}
\begin{tabular}{lccc}
\hline
Name & RA &  DEC  & Area  \\
 & deg & deg & deg$^2$  \\
\hline
D33 & 100 to 280 to  & +32.5 to +34.5 & 300 \\
S82a & 36 to 42 & $-$1.3 to +1.3 & 13 \\
S82b & 350 to 10 & $-$1.3 to +1.3 & 46 \\
S82c & 317 to 330 & $-$1.3 to +1.3 & 30 \\
COSMOS & 330 to 336 & $-$1.3 to +1.3 & 10 \\
\hline
\end{tabular}
\label{tab:overlap}
\end{table}

\begin{table}[ht]
\centering
\caption{Observing Schedule}
\label{tab:observing}
\begin{tabular}{cccccc}
  \hline
  Survey & Telescope/Instrument & Nights & Start & Finish  & Bands \\
  \hline
  DECaLS & Blanco/DECam & 145 & 2014 Aug & 2019 Mar & $g$,$r$,$z$ \\
  BASS & Bok/90prime & 250 & 2015 Jan & 2019 Mar & $g$,$r$ \\
  MzLS & Mayall/Mosaic-3 &  383 & 2016 Feb & 2018 Feb & $z$ \\
  \hline
 \end{tabular}
 \end{table}

\begin{table}[!th]
\begin{center}
\caption{Depths and Delivered Image Quality}
\begin{tabular}{lccccccccc}
\hline
Survey & \multicolumn{6}{c}{Single-Frame Depths$^1$} & \multicolumn{3}{c}{DIQ$^4$} \\
Name   & \multicolumn{3}{c}{PSF Depth$^2$} & \multicolumn{3}{c}{Galaxy Depth$^3$} & \multicolumn{3}{c}{($''$)} \\
       & $g$ & $r$ & $z$ & $g$ & $r$ & $z$ & $g$ & $r$ & $z$ \\
\hline
DECaLS$^5$ & 23.95 & 23.54 & 22.50 & 23.72 & 23.27 & 22.22 & 1.29 & 1.18 & 1.11 \\
BASS$^6$  & 23.65 & 23.08 & & 23.48 & 22.87 & & 1.61 & 1.47 & \\
MzLS$^7$  &  & & 22.60 & & & 22.29 &  &  & 1.01 \\
\hline
\end{tabular}
\end{center}
$^1$ In AB mag. \\
$^2$ Median 5$\sigma$ detection limit in AB mag for a point source in individual images. \\
$^3$ Median 5$\sigma$ detection limit in AB mag for the fiducial DESI target (galaxy with an exponential disk profile with $r_{\rm half}=0.45\arcsec$). \\
$^4$ Delivered image quality, defined as the FWHM in arcseconds of the measured point spread function. For comparison, the corresponding median FWHM for the SDSS imaging is $\approx0.85\times{\rm\tt psfWidth_{\rm SDSS}}$ = 1.22, 1.12, 1.10 arcsec in the $g$, $r$, and $z$ bands, respectively (see \url{https://www.sdss.org/dr14/imaging/other_info/\#SeeingandSkyBrightness}).)\\
$^5$ From Data Release 5. \\
$^6$ From Data Release 6. \\
$^7$ Based on all data obtained for the survey.
\label{tab:depths}
\end{table}

\subsection{DECaLS: The Dark Energy Camera Legacy Survey}
\label{sec:DECaLS}

The Dark Energy Camera \citep[DECam;][]{decam} at the 4-m Blanco telescope at the Cerro Tololo Inter-American Observatory is the most efficient imager for wide-field surveys currently available. DECam has 62\footnote{One CCD died before the survey, one is only partially usable, and one was inoperative for part of the survey.} 2048x4096 pixel format 250$\mu$m-thick LBNL CCDs arranged in a roughly hexagonal $\approx$3.2~deg$^2$ field of view. The pixel scale is $\approx$0.262~arcsec/pix. In addition to the wide field of view, DECam provides high sensitivity across a broad wavelength range ($\sim$400--1000\,nm) and low operational overheads. We are therefore conducting the bulk of the imaging for the Legacy Surveys with DECam. DECam is already being used by the Dark Energy Survey \citep[DES;][]{des} to cover $\approx$5000~deg$^2$ in the SGC, $\approx$1130~deg$^2$ of which lie within the DESI footprint.  The Dark Energy Camera Legacy Survey (DECaLS) is targeting the remaining $\approx$9350~deg$^2$  ($\approx$3580~deg$^2$ in the SGC and $\approx$5770~deg$^2$ in the NGC). DECaLS was the first of the three Legacy Surveys to begin observations (in August, 2014) and therefore defined the $grz$ bandpasses and strategy for the other two surveys described in this section.

For the DECaLS observations we adopt a tiling pattern (from Hardin, Sloane and Smith\footnote{\url{http://neilsloane.com/icosahedral.codes/}}) which can cover the entire sky with 15,872 tiles and which results in an effective area per tile of 2.60~deg$^2$. 
In order to fill gaps between the CCDs and achieve the required depth across the maximum area, we have chosen three similar, but offset, tiling patterns (labeled Pass 1, Pass 2 and Pass 3).  Pass 2 is offset by ($\Delta\alpha$, $\Delta\delta$) = ($0.2917^\circ, 0.0833^\circ$) deg relative to Pass 1; Pass 3 is offset by ($0.5861^\circ, 0.1333^\circ$). 
When the survey is complete, approximately 99.97\%, 98.00\%, 74.33\% and 23.8\% of the survey will have, respectively, at least 1, 2, 3 and 4 exposure coverage.


DECam can reach the required depths for the fiducial DESI target (see \S~\ref{sec:desirequirements}) in total exposure times of 140, 100 and 200 sec in $g$, $r$, $z$ in ``nominal'' conditions, defined as photometric and clear skies with seeing FWHM of 1.3 arcsec, airmass of 1.0 (i.e., zenith pointing), and sky brightness of 22.04, 20.91, and 18.46 AB mag~arscsec$^{-2}$, respectively. Accounting for weather loss, DECam is capable of imaging 9000 deg$^2$ of the footprint of the Legacy Surveys to this depth in 157 scheduled nights. Observations in the $g$ and $r$-band filters are only obtained during dark periods when the moon is below the horizon; $z$-band observations are obtained when the moon is in the sky and during the morning and evening twilight. The DECam observations are conducted using a dynamic observing mode, where the exposure times and target field selection are modified on-the-fly based on the observing conditions to ensure uniform depth to the extent possible (see \S~\ref{sec:dynamic} for details). The median FWHM of the delivered image quality (DIQ)is $\approx 1.3$, 1.2, and 1.1 arcseconds in the $g$, $r$ and $z$ bands respectively for the DECaLS survey.
 
``The DECam Legacy Survey of the SDSS Equatorial
Sky'' (NOAO Proposal ID \# 2014B-0404; PI: D.\ Schlegel and A.\ Dey), was
initially proposed as a public survey beginning in semester 2014A
as part of the NOAO Large Surveys programs.  This project was initially allocated 64 nights and was aimed at imaging the existing SDSS footprint at $\delta\le +32^\circ$. 
The imaging program has been supplemented to a total of 157 scheduled
nights (first by NOAO Proposal ID \# 2016A-0190, and later using a Director's allocation) to 
enlarge
the footprint to the full DESI equatorial footprint (i.e., the full region labeled DECaLS in Figure\,\ref{fig:zsfootprint}).
The goal is to complete this survey in the 2019A semester.


The Legacy Surveys program also makes use of other DECam $grz$ data
within the DESI footprint, as those data become public.
The most significant of these other data sets is from
the Dark Energy Survey, which includes a 1,130~deg$^2$
contiguous area in the SGC footprint of the Legacy Surveys. 
DECaLS is therefore {\em not} re-observing that area,
and is instead making use of the DES raw data as they become public.
Data from the early DECam science verification period have a number
of problematic features, and are not currently included in the reductions or data releases from the Legacy Surveys.

\subsection{BASS: The Beijing-Arizona Sky Survey}
\label{sec:BASS}

The Beijing-Arizona Sky Survey \citep[BASS;][]{bassOverview} is imaging the 
DEC~$\ge +32^\circ$ region of the DESI North Galactic Cap footprint ($\approx$5,100~deg$^2$) 
in the $g$ and $r$ optical bands. BASS
uses the 90Prime camera \citep{90Prime} at the prime focus of the Bok 2.3-m telescope.
The Bok Telescope, owned and operated by the University
of Arizona, is located on Kitt Peak,
adjacent to the Mayall Telescope. 
The 90Prime instrument is a prime focus 8k$\times$8k
CCD imager, with four University of Arizona ITL 4k$\times$4k CCDs
that have been thinned and UV optimized with peak QE of 95\% at
4000\,\AA\ \citep[see][for details]{90Prime}.  These CCDs were installed
in 2009 and have been operating routinely since then.  90Prime
delivers a $1.12^\circ$ field of view, with 0.45$''$ pixels, and 94\%
filling factor. The median FWHM of the delivered image quality at the telescope
is 1.6$''$ and 1.5$''$ in the $g$- and $r$-bands, respectively.  
The throughput and performance
in these bands were demonstrated with data in September, 2013.

BASS tiles the sky in three passes, similar to
the DECaLS survey strategy.  At least one of these passes
is observed in photometric conditions (Pass 1) and seeing
conditions better than 1.7\arcsec. Observations in $g$-band are restricted to dark time, when the moon is below the horizon. The typical individual exposure times are 100~sec per band, with the requirement that 3 passes are needed to reach depth. As in the case of DECaLS, the exposure times are varied depending on the conditions, but limited between 50~sec and 250~sec. We refer the reader to \citep{bassOverview} for further details.

BASS was awarded 56/100/100/90 nights in the 2015A/2016A/2017A/2018A semesters 
(PIs: Zhou Xu and Xiaohui Fan)
to target 5500~deg$^2$ in the NGC and $\approx$100~deg$^2$ in the SGC.\footnote{see {\tt http://batc.bao.ac.cn/BASS}}  
These areas include $\approx$400~deg$^2$ of overlap 
with regions covered by other components of the Legacy Surveys (Table~\ref{tab:overlap}) in order to cross-calibrate photometry.
Prior to the start of BASS it was determined that the existing Bok $g$-band filter was well-matched to the DECam
$g$-band filter but the existing Bok $r$-band filter had a significantly different
bandpass. A new $r$-band filter was therefore acquired from Asahi in April 2015, and was used for 
subsequent BASS observations.
The 90Prime camera has excellent response at blue wavelengths, and as a result the effective throughput as a function
of wavelength for the $g$ and $r$ photometric bands in the BASS survey is different than that for the same
bands in the DECaLS survey.

The BASS survey began observations in Spring 2015. A number of instrument
control software updates, new flexure maps, and new observing tools
were implemented that greatly improved the pointing accuracy,
focusing of the telescope, and observing efficiency. A total of
15\% of the $g$-band and 2\% of the $r$-band tiles were observed
in Spring 2015.  It was discovered that those data suffered
from defective electronics in the read-out system that introduced
analog-to-digital conversion errors, gain variations and non-linearities.  The 90Prime CCD controller electronics were replaced in September 2015 followed by a recommissioning of
the system in Fall 2015.

BASS completed 40\% of its expected coverage in 100 scheduled nights
in the 2016A semester (January--July). 
BASS is expected to complete observations by March 2019. As of December 2018, the BASS project has undergone two data releases that 
are detailed in \citet{bassDR1,bassDR2}.


\subsection{MzLS: The Mayall {\it z}-band Legacy Survey}
\label{sec:MzLS}

The Mayall $z$-band Legacy Survey (MzLS) has imaged the 
$\delta\ge +32^\circ$ region of the NGC footprint of the Legacy Surveys. These $z$-band observations complemented the BASS $g$ and $r$ band observations in the same $\approx$5,100~deg$^2$ sub-region of the Legacy Surveys. The delivered image quality at the Mayall telescope is significantly better than that at the Bok telescope (median of $\approx1.0\arcsec$ vs $\approx1.6\arcsec$) and hence the MzLS data are critical to deblending images and to deriving morphologies and source models for the photometric catalogs. 

MzLS used the Mosaic-3 camera at the prime focus of the 4-meter
Mayall telescope at Kitt Peak National Observatory.
In 2015, prior to the commencement of MzLS, the Mayall 4-m telescope's prime focus imaging system underwent a major upgrade aimed at improving its $z$-band efficiency. Details of the Mosaic-3 camera upgrade are presented in \citet{mosaic3}; here, we briefly describe the main modifications to the system. 


The Mosaic-3 camera is a new version of the prime focus imaging system at the Mayall 4-m telescope. The previous version, known as Mosaic-1.1, was a blue-sensitive camera equipped with eight thinned 2048$\times$4096 15$\mu$m pixel format e2v CCDs. The camera had a twin, Mosaic-2, at the Blanco telescope at CTIO, which was decommissioned and replaced with the Dark Energy Camera. The Mosaic-3 upgrade repurposes the dewar from the CTIO Mosaic-2 camera, while retaining the rest of the Mosaic-1.1 mechanical system and guider. Yale University designed and built a new cold plate for the dewar, which was populated with four (500$\mu$m-thick) fully-depleted LBNL 4096$\times$4096 15$\mu$m pixel CCDs. The new readout system consists of four prototype DESI controllers, one for each CCD, that are synchronized to a single clock in order to simultaneously read the four quadrants of each device. The dewar was delivered to NOAO in September 2015 where it was integrated with the Mosaic-1.1 mechanical enclosure, shutter, filter wheel and acquisition and guider system. NOAO also purchased a new $z$-band filter, matched to the DECam filter bandpass, in order to minimize any differences between the DECam and Mosaic-3 $z$ surveys. In addition, the KPNO 4-m telescope control system and the imaging camera software were upgraded for improved operational efficiency \citep{abareshi2016,mosaic3}. Mosaic-3 saw first light in October 2015 and underwent further on-sky commissioning runs in 
November and December 2015.  The $z$-band efficiency with Mosaic-3 is
measured to be 60\% better than that of its predecessor, the Mosaic-1.1 camera.

The MzLS survey uses a 3-pass strategy, similar to DECaLS, and tiles the sky with $\approx$ 122,765 tiles per pass.  Pass 1 is observed only in photometric conditions and seeing conditions better than 1.3 arcsec. For 1.3 arcsec seeing and a sky brightness of 18.2 AB mag/arcsec$^2$,
the total time required is 200 sec ($\approx67$~sec per exposure) in
$z$. As in the case of DECaLS, we limited the exposure times for individual exposures to be in the range $80\le t_{\rm exp}\le 250$~sec. Observations were made during all lunar phases, although during bright time we limited our observations to regions of the footprint lying $>$40--50~deg away from the Moon. 

MzLS began official survey operations on February 2, 2016, and ended on February 12, 2018. During this period, MzLS used a total of 382.7 nights, 276.8 of which were clear enough to allow observations. During the second semester of observing (2017A), MzLS progress slowed because of poor weather and instrumental and operational problems. 

The Mosaic-3 camera was decommissioned and the Mayall telescope shut down on February 12, 2018 to prepare for the installation of the DESI instrument.


\section{{\it WISE} Data}
\label{sec:WISE}

The Legacy Surveys source catalogs include mid-infrared photometry from the {\it Wide-field Infrared Survey Explorer} ({\it WISE}) satellite for all optically detected sources. 
Mid-infrared imaging is critical to the DESI targeting algorithms for luminous red galaxies (LRGs) and quasars (QSOs).  During its primary 7-month mission from 2010 January through 2010 August, {\it WISE} conducted an all-sky survey in four bands centered at 3.4, 4.6, 12 and 22\,$\mu$m \cite[known as W1, W2, W3 and W4;][]{WISE, cutri12}. Following its primary 4-band mission, {\it WISE} continued survey operations in the three bluest bands for 2 months, then the two bluest bands for an additional 4 months, resulting in a combined 13-month mission that completed in 2011 February. Through a mission extension referred to as {\it NEOWISE-Reactivation} \citep[{\it NEOWISE-R};][]{neowiser}, NASA reactivated the satellite and resumed 2-band survey observations on 2013 December 13. {\it NEOWISE-R} observations remain ongoing. 
Annual {\it NEOWISE-R} data releases, each consisting of single-exposure (Level 1b) images and source extractions, have occurred on 2015 March 25, 2016 March 23, 2017 June 1 and 2018 April 19.



DESI target selection utilizes the two shortest-wavelength bands
at 3.4\,$\mu$m (W1) and 4.6\,$\mu$m (W2).  Photometry in these bands is measured using \tractor\ algorithm (see Section~\ref{sec:tractor}), adopting source centroid and morphology parameters from the optical imaging, which has much better angular resolution than {\it WISE}. \tractor\ measurements are based on custom stacks of {\it WISE/NEOWISE} exposures which are optimized for forced photometry and therefore preserve the native {\it WISE} resolution. These stacks are referred to as unWISE coadds \citep{Lang2014}. DR1 made use of the \citet{Lang2014} unWISE coadds based on the initial 13-month {\it WISE} data set, reaching 5$\sigma$ limiting magnitudes of 20.0 and 
19.3\,AB mag in W1 and W2. 
Subsequent Legacy Surveys releases have used a series of updated, deeper unWISE coadd data sets featuring progressively more {\it NEOWISE-R} imaging 
\citep[][see Table 6]{fulldepth_neo2,fulldepth_neo1}. DR7 incorporates all five years of publicly available {\it WISE} and {\it NEOWISE-R} imaging, including that from the fourth-year {\it NEOWISE-R} release. 
The final catalogs from the Legacy Surveys will push even deeper at $3$--$5\,\mu$m by leveraging the full {\it WISE} and {\it NEOWISE-R} data sets.


In addition to the mid-infrared photometry measured from the ``full-depth'' W1/W2 unWISE stacks (which are required for DESI targeting), the Legacy Surveys DR3--DR7 also include W1/W2 forced photometry light curves corresponding to all optically detected sources. These light curves are measured from time-resolved unWISE coadds similar to those described in \cite{unwise_time_resolved2,unwise_time_resolved3}. Such light curves provide variability information on all optically-detected sources, which can be used, among other things, for the DESI quasar selection, although this possibility has not yet been tested in detail. 
In DR7, the Legacy Surveys W1/W2 light curves typically have 10 coadded epochs per band, spanning a $\approx$7.5 year time baseline.

\section{Observations \label{sec:observing}}

In this section, we briefly describe the observing strategy employed by the Legacy Surveys. For a more detailed description of the implementation and algorithms, we refer the reader to \citet{LSstrategy}.

\subsection{Survey Strategy}

As described in \S\,\ref{sec:surveys}, all three surveys (DECaLS, BASS and MzLS) use a 3-pass strategy to tile the sky. This strategy is designed to account for the gaps between CCDs in the cameras, ensure that the surveys reach the required depth, remove particle events and other systematics, and ensure photometric and image quality uniformity across the entire survey. In order to calibrate the entire survey photometrically, we place requirements on the minimum observing conditions needed for each pass. 
Pass 1 tiles are only observed when the conditions are photometric (defined as the transparency being better than 90\% and the sky being clear) and when the seeing is better than a specified limit (1.3\arcsec\ for DECaLS and MzLS; 1.7\arcsec\ for BASS). If only one of these conditions is met (i.e., seeing $<$ 1.3\arcsec/1.7\arcsec\ or photometric), then we observe pass 2; if both are not met, we observe pass 3. The successful implementation of this strategy guarantees that we have at least one photometric and good-seeing image at every sky position, which can be used to calibrate the photometry across the entire survey footprint. 

The determination of whether the conditions are photometric and the seeing measurements are made ``on-the-fly'' at the telescope, using a combination of the on-site telemetry, the observer's periodic visual inspection of the sky, and quick analyses of every frame. At the Blanco telescope, the observers determine which pass to observe using the output of the Radiometric All-Sky Infrared Camera \citep[RASICAM;][]{rasicam}, the CTIO All-Sky Camera\footnote{\url{http://www.ctio.noao.edu/noao/content/tasca-latest-image}}, the output of the DECam ``kentools" (created by S.\ Kent) and our own custom software. Our software identifies stars, matches to the PS1 Data Release 1 (DR1) catalog, and measures the seeing, transparency, sky brightness and positional offset of the telescope from the desired pointing center. At the Mayall and Bok telescopes, the observers determine which pass to observe using the KPNO All-Sky Camera, weather satellite maps, and 
our own custom \href{https://github.com/legacysurvey/obsbot}{software}\footnote{\url{https://github.com/legacysurvey/obsbot}. }. 

\subsection{Dynamic Observing \label{sec:dynamic}}

In order to optimize the observing efficiency and create as uniform a survey as possible, we have implemented an observing mode which adjusts the exposure time and optimizes the selection of target fields for observation automatically based on the observing conditions. The observing strategy is described in detail in \citet{LSstrategy}, but here we provide a brief overview. 

The desired target depth of each exposure is defined as that for which the fiducial DESI target galaxy (see \S~\ref{sec:desirequirements}) is detected with a signal-to-noise ratio of at least $5/\sqrt{2}$ (i.e., that the survey reaches the requisite depth with two passes). To ensure that each image of the sky reaches the desired depth, we implement the following procedure. We plan image exposure times based on knowledge of the target field's interstellar dust reddening, its position on the sky at the likely time of observation (which determines the likely atmospheric extinction, sky brightness, and modulates the seeing), and estimates of the observing conditions. 
As soon as an image is taken and written to disk, a sample CCD (or single amplifier of a CCD) is analyzed: sources are detected and their positions are cross-matched with a stellar catalog derived from the PS1 survey. This analysis produces estimates of the seeing, transparency (estimated by comparing the measured zero point with the fiducial photometric zero point of an observation through clear skies), the telescope pointing error, the sky brightness and the resulting depth reached for the canonical DESI galaxy target. These measurements allow us to update the exposure time of subsequent observations to ensure that we reach the required depth. 
We scale exposure times by a factor $f=T^{-2} 10^{0.8k_i(X-1)} 10^{0.8A_iE_{\rm B-V}} 10^{-0.4(\Delta m_{\rm sky})} (N_{\rm eff}/N_{\rm eff,fid})$, where $T$ is the transparency, $k_i$ is the atmospheric extinction coefficient for band $i$, $X$ is the airmass, $A_i$ is the Galactic dust extinction coefficient for band $i$, $E_{\rm B-V}$ is the Galactic dust reddening along the line of sight, $\Delta~m_{\rm sky}$ is the difference in the sky brightness from the fiducial (i.e., 22.04, 20.91, 18.46 AB mag arcsec$^{-2}$ in $g$, $r$, $z$, respectively), and $(N_{\rm eff}/N_{\rm eff,fid})$ is a measure of the PSF area (in pixels) relative to the fiducial. 
Exposure times are not allowed to fall below a minimum value in order to limit the overhead\footnote{For example,  for DECam, these were initially defined as 50, 50, and 100\,sec for $g$,$r$,$z$ respectively; after 2016-07-20, the minimum exposure time in $g$ was increased to 70~sec.}.   Additionally, exposure times are limited to a maximum value defined by the minimum of $t_{\rm sky}$ and $t_{\rm max}$, where $t_{\rm sky}$ is the exposure time at which the sky counts = 20,000\,adu, and $t_{\rm max}$ is a fixed maximum exposure time (e.g., $t_{\rm max}$ is [200,175,250]\,sec for DECam [$g$,$r$,$z$] observations, respectively). 

In practice, it takes a minimum lag of two exposures to update the queue with an observation that has a modified exposure time. At the Blanco, this lag was driven by the need to keep at least two exposures in the active queue to avoid stopping the queue inadvertently, and at the Mayall the transfer of images and the subsequent analyses resulted in this lag. Even with the current implementation, the result is a relatively uniform survey product \citep[see][for details about the algorithms used and the implementation]{LSstrategy}.

\section{Data Reduction and Calibrations \label{sec:redux}}

All data from the Legacy Surveys are first processed at NOAO/Tucson through the NOAO Community Pipelines (``CPs''). Each instrument and telescope combination has its own CP that takes raw data as an input and provides detrended and calibrated data products. The NOAO Pipeline Scientist and architect is co-author F. Valdes, who is responsible for the development and continued operation of the CPs. The CPs include algorithms and code (from a variety of sources, the key ones being code developed by DES Data Management, the TERAPIX suite, and IRAF; \citet{desCP,sextractor,psfex,terapix,iraf}) which are modified and packaged for the needs of the NOAO environment and characteristics of the different instruments.  A common feature of all the CPs is the orchestration framework (The NOAO High Performance Pipeline System: \cite{NHPPS}) that allows 
parallelized processing across the NOAO computing resources to handle the large volumes of data produced by NOAO observing programs. 

The CPs provide instrumentally calibrated data products for observers, programs, and archival researchers.  Instrumental calibrations include: typical CCD corrections (e.g., bias subtraction and flat fielding); astrometric calibration (e.g., mapping the distortions and providing a world coordinate system, or WCS); photometric characterization (e.g., magnitude zero point calibration); and artifact identification, masking and/or removal (e.g., removal of cross-talk and pupil ghosts, and identification and masking of cosmic rays).  Data products delivered to the NOAO Science Archive include flux calibrated images (i.e., individual images with and without distortion corrections applied, and image stacks), bad data masks, and weight maps.  

The three cameras used by the Legacy Surveys (i.e., DECam, Mosaic-3, and 90Prime) each have their own CP.  The basic steps of each CP are summarized in Table~\ref{tab:CPsteps}.  Detailed technical descriptions of each CP are in preparation\footnote{A draft on-line DECam CP  description is \url{https://www.noao.edu/noao/staff/fvaldes/CPDocPrelim/PL201_3.html}} \citep[][describes an early version of DECam CP]{desCP}. The CP for the DECaLS data evolved from the Dark Energy Survey pipeline such that it has algorithms and code from several sources.  The key sources are code developed by DES Data Management, the TERAPIX suite, and IRAF \citep{desCP,sextractor,psfex,terapix,iraf}.
Some calibrations are not perfect, with the detection and masking of artifacts being only partially effective and background pattern subtraction around very large and bright sources being prime examples. In particular, the CP can result in unmasked spurious sources in the final catalogs. First, the thick, deep depletion LBNL CCDs employed in the DECam and Mosaic-3 cameras are excellent detectors of particle events \citep[see][for a more detailed discussion]{groom2004}, a fraction of which are inadequately masked by the current CP. Second, asteroids and other moving targets are not flagged by the CP and may appear as detected sources in the catalogs (at least through DR7).

The CP-calibrated individual images, bad pixel masks and weight maps are transferred to the National Energy Research Scientific Computing Center (NERSC), where post-processing is done in order to improve the astrometric and photometric calibrations and create the source catalogs. Similarly, the {\it WISE} satellite data are transferred to NERSC as they become public, and new coadded stacks are constructed on an approximately yearly basis. 

\begin{table}[!th]
\begin{center}
\caption{Calibration Steps in the NOAO Community Pipelines}
\begin{tabular}{lccc}
\hline
{\bf \# \ Calibration} &  {\bf DECam}  &  {\bf BASS}  & {\bf MzLS} \\
\hline
1. \ Linearity correction   & \checkmark  &            &  \\
2. \ Cross-talk subtraction &    \checkmark & \checkmark & \checkmark \\
3. \ Overscan \& bias subtraction   &    \checkmark & \checkmark & \checkmark \\
4. \ Dome flat fielding  &    \checkmark & \checkmark & \checkmark \\
5. \ Amplifier gain balancing\tablenotemark{1}  & \checkmark  & \checkmark  & \checkmark \\
6. \ Masking of bad pixels\tablenotemark{2} & \checkmark & \checkmark & \checkmark \\
7. \ Interpolation over bad/saturated pixels\tablenotemark{3} & \checkmark & \checkmark & \checkmark \\
8. \ Correction of line shifts\tablenotemark{4}  &     & \checkmark & \checkmark \\
9. \ Astrometric calibration\tablenotemark{5,6}  & \checkmark & \checkmark & \checkmark \\
10. Removal of sky patterns/gradients & \checkmark & \checkmark & \checkmark \\
11. Pupil ghost subtraction & \checkmark & only $g$-band & \checkmark \\
12. Fringe-pattern removal\tablenotemark{7} & only $z$-band & only $r$-band & \checkmark \\
13. Illumination correction (sky flat) & \checkmark &  &  \\
14. Removal of pattern/striping noise &         & \checkmark & \checkmark \\
\hline
\end{tabular}
\end{center}
\tablenotetext{1}{DECam uses starflats and BASS/MzLS uses PS1. For MzLS the gain balancing is a function of the sky level.}
\tablenotetext{2}{Bad pixels are detector defects, saturated, bleed trails, cosmic rays, and satellite trails.}
\tablenotetext{3}{Stellar cores are masked but not interpolated.}
\tablenotetext{4}{Some MzLS data suffered from 1/3-pixel shifts and dropped columns and BASS has systematic centroid shift due to CTE.}
\tablenotetext{5}{DECam is referenced to a mixture of 2MASS and Gaia DR1.  BASS and MzLS are referenced only to Gaia DR1. }
\tablenotetext{6}{DECam has a fixed distortion map with 2nd order adjustments. BASS and MzLS have full 4th order solutions.}
\tablenotetext{7}{Implemented for DECam only in 2018, and so far only applied to the 2018 observations.}
\label{tab:CPsteps}
\end{table}

\subsection{Astrometric Calibration}

The NOAO CP reductions of all Legacy Survey imaging data derive a world coordinate system (WCS), a function mapping pixel coordinates to celestial coordinates.  The function (TPV: tangent plane projection with polynomial distortions\footnote{ \url{https://fits.gsfc.nasa.gov/registry/tpvwcs/tpv.html}}) is determined for each CCD by least square fitting to the pixel centroids of detected sources with known coordinates in a reference catalog. For the astrometric solution, the pixel centroids of reference stars are computed using intensity weighted means using Source Extractor \citep{sextractor} in the DECam CP, and using the ACE package in IRAF \citep{ace,iraf} in the Mosaic-3 and 90Prime CPs. The final source positions in the catalogs are computed using the Tractor, as described in section 8. 

The reference coordinates were from the 2MASS catalog \citep{2MASS} for DECam data from 2013 and 2014; later DECam data and all Mosaic-3 and 90Prime data use the Gaia DR1 catalog \citep{gaiaDR1}.  Since the calibration procedure ties source positions in each exposure to celestial sources, it effectively includes calibration for the atmospheric distortions (at some mean color) and, in the case of the Bok 90Prime data, correction for distortions resulting from the focusing procedures.

While this procedure corrects the data for global distortions (the TPV solutions are continuous and smooth across an individual CCD), it does not correct for some small-scale effects. For example, the DECam and Mosaic-3 CCDs are known to have very small scale distortions known as ``tree rings'' \citep{treeRings}, and the Mosaic-3 CCDs show a residual astrometric pattern from bonding stresses. Differential chromatic refraction is also not accounted for in the astrometric solutions \citep[e.g., see][for an excellent discussion of all the issues affecting the DECam astrometry]{decamastrom}. These combined effects affect the astrometric accuracy on individual CCDs at the level of $\approx10-30$~mas and are currently not corrected in the post-processing catalog generation step. 

%

In addition, the Mosaic-3 electronics occasionally read out with a missing starting column. The CP detects and corrects this; however, since the edges are masked anyway this has no effect on the astrometry. Prior to MJD 57674 the readout electronics introduced a one-third pixel shift between amplifiers in the vertical transfer direction (corresponding to the east-west direction on the sky).  Since this is a precise, discrete offset the CP corrects this completely with no effect on the astrometry. 

90Prime has charge transfer effects that affect centroid measurement and, in particular, introduce systematic opposing shifts between amplifiers in the serial transfer direction. This effect is evident as a discrete jump at the amplifier boundaries in the astrometric offset when comparing the astrometry of reference stars from the Gaia DR1 catalog with that measured from the Bok data using the best-fit smooth astrometric solution. 
The systematic offset between the two halves is $\approx160$\,mas for CCD-1 and $\approx70$\,mas for the other 3 CCDs. The CP applies a relative shift to the pixels from each amplifier so that the astrometric offset jump across the boundary is minimized. These corrections are applied to each exposure and they substantially reduce, but do not completely remove, this systematic effect. The residual difference does show small temporal variations (of order a few mas) from night-to-night; correcting for this residual would require a higher order correction to the astrometry (rather than just a zero point offset at the boundary), which may result in changing the shape of the PSF. 

\begin{figure}[!th]
\centering\includegraphics[width=0.65\textwidth]{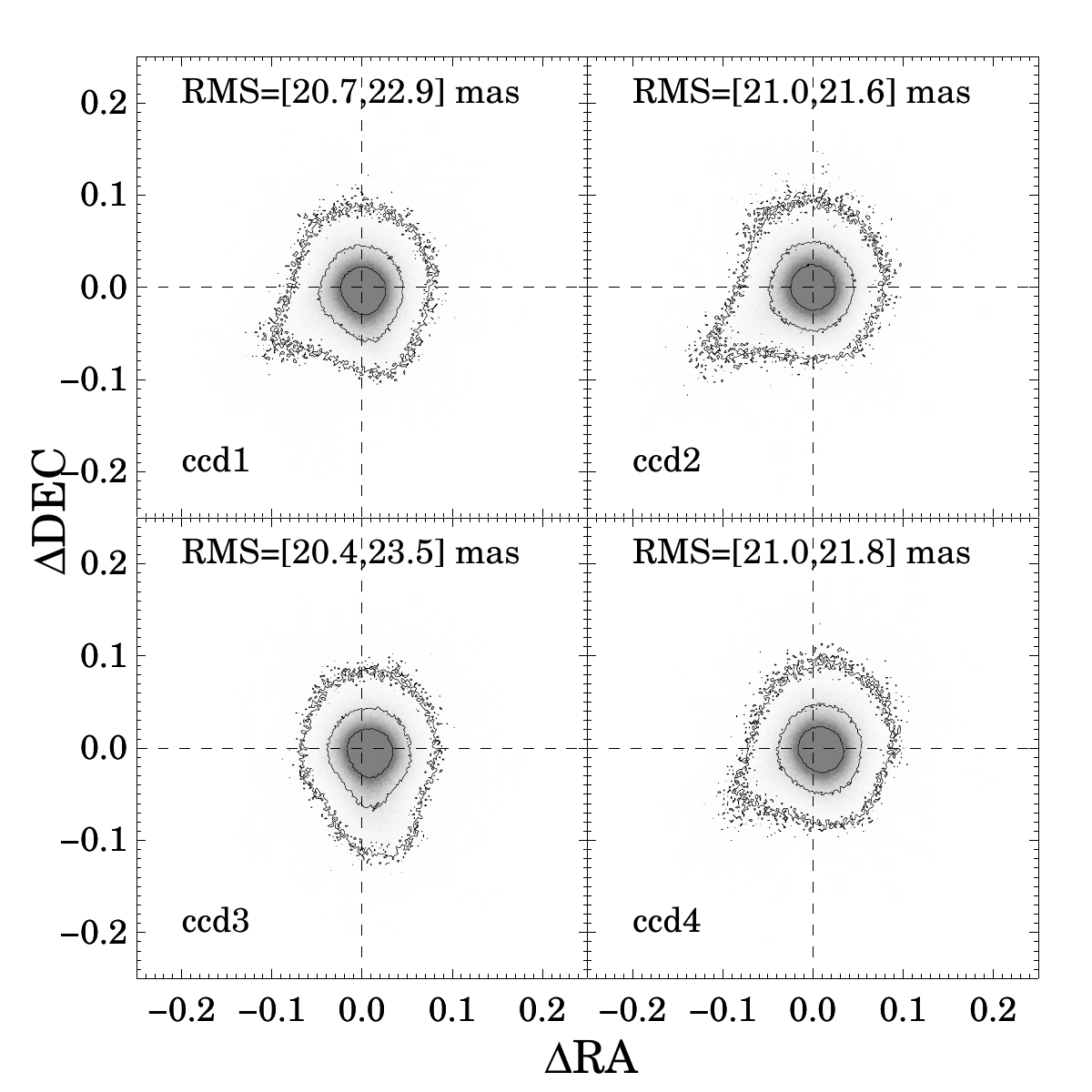}
\caption{The astrometric precision of the four Mosaic-3 CCDs, computed by matching stars detected in the MzLS images with those in Gaia DR1 catalog. In each panel, the greyscales show the distribution of the differences (in units of arcseconds) between the derived positions (using the WCS) of the centroids of bright stars on a Mosaic-3 CCD and their positions in the Gaia DR1 catalog. \label{fig:astrometryMzLS}}
\end{figure}
\begin{figure}[!th]
\centering\includegraphics[width=0.65\textwidth]{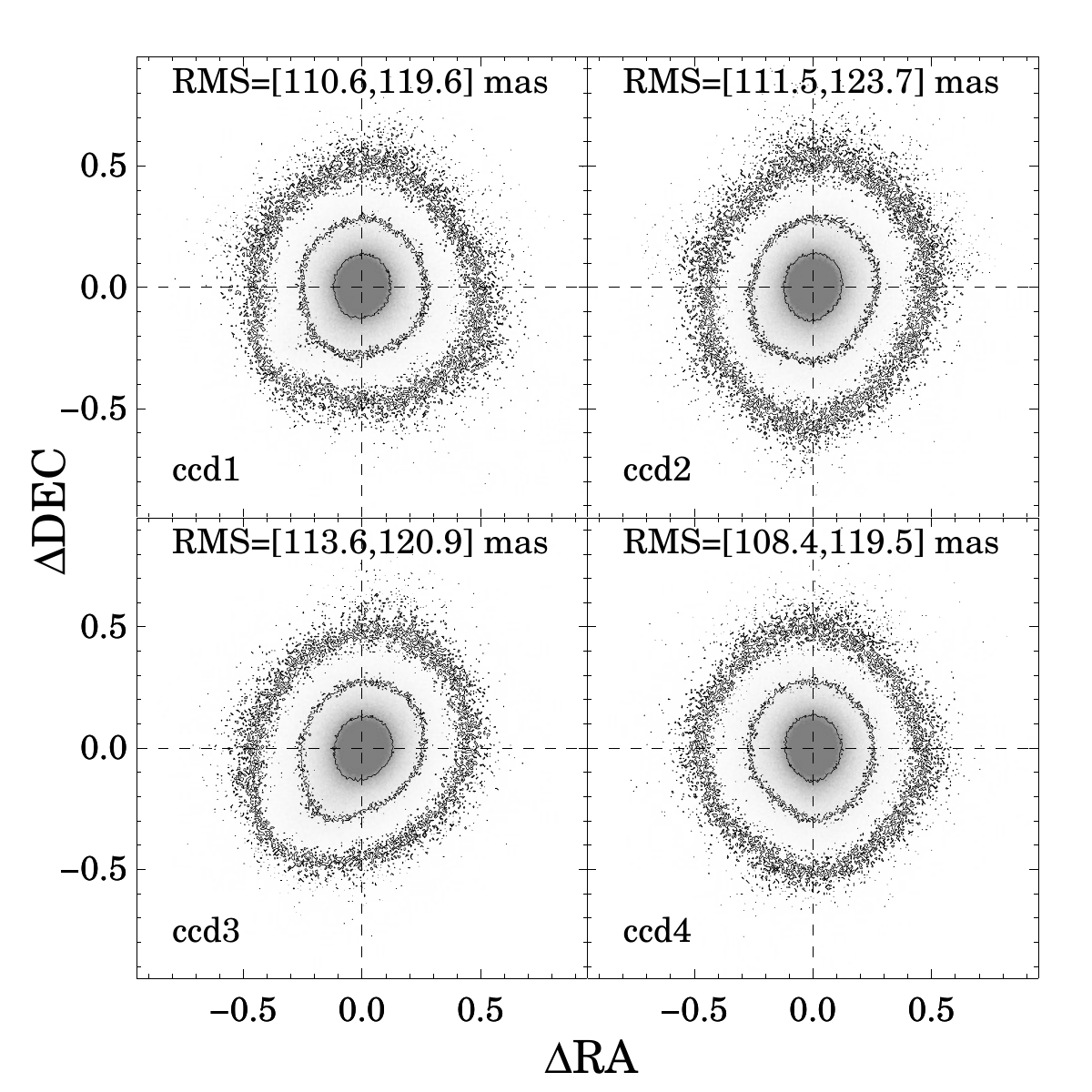}
\caption{The astrometric precision of the four Bok 90Prime CCDs, computed by comparing the derived positions (using the WCS) of bright stars with their positions in the Gaia DR1  catalog. \label{fig:astrometryBASS}}
\end{figure}
\begin{figure}[!th]
\centering\includegraphics[width=0.75\textwidth]{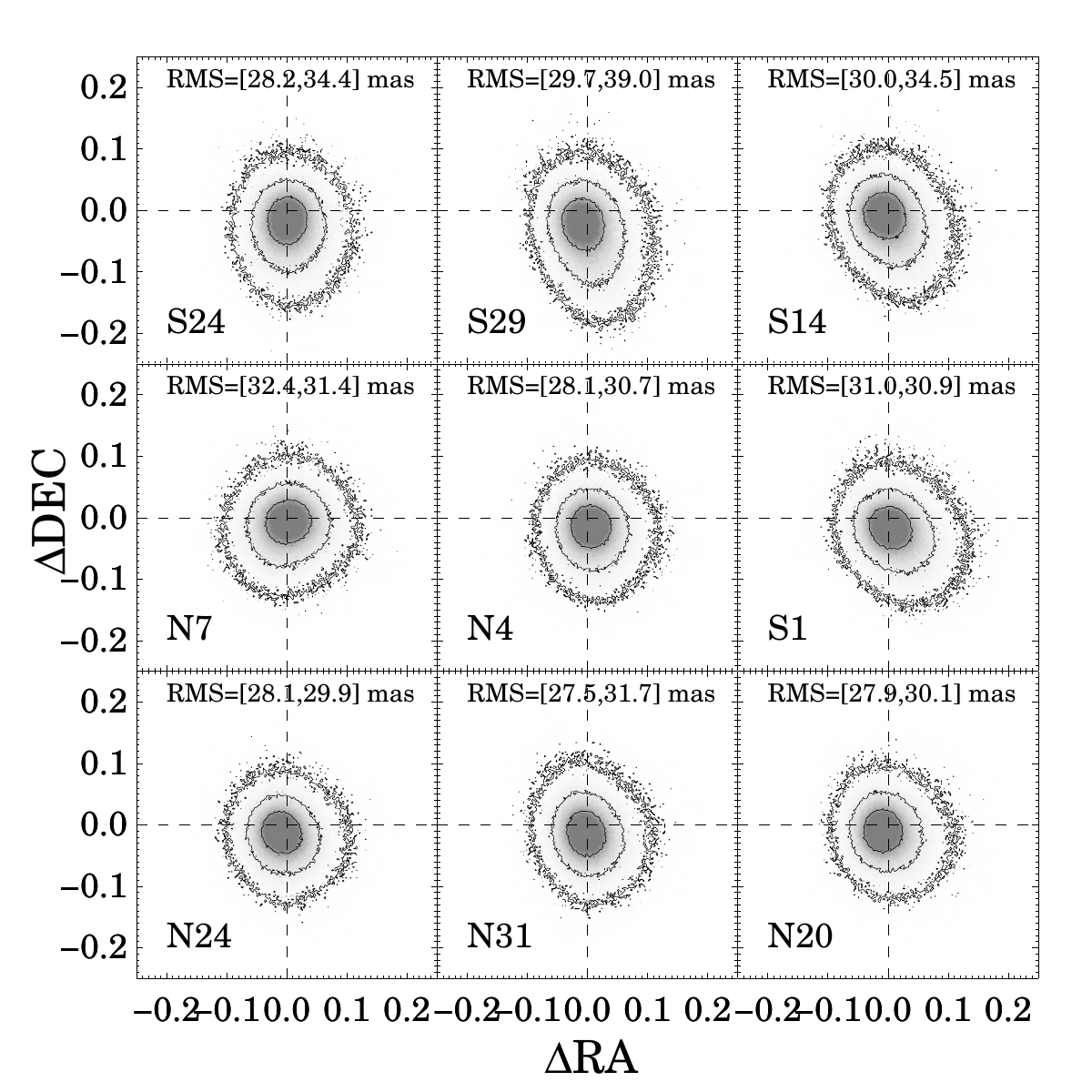}
\caption{The astrometric precision of the DECaLS CCDs, computed by comparing the derived positions (using the WCS) of bright stars with their positions in the Gaia DR1 catalog. The N4 CCD is one of two central CCDs in the DECam mosaic; the other 8 CCDs shown are edge CCDs in the mosaic and represent regions with the largest astrometric and PSF distortions. \label{fig:astrometryDECaLS}}
\end{figure}

As noted earlier, the CP data is only the first part of the Legacy Surveys calibrations.  Small residual mean offsets per CCD are applied to the CP astrometric zero point calibrations using a reference catalog constructed from stars color-selected from the PS1 DR1 catalog \citep{panstarrs} with positions from the Gaia DR1 catalog \citep{gaiaDR1}.  For all releases prior to (and including) DR3 of the Legacy Surveys, astrometric and photometric calibration is based on comparisons to a subset of PS1 catalog sources with  magnitudes $<$ 21.5\,AB\,mag and colors $0.4 < (g-i)_{\rm PS1} < 2.7$\,AB\,mag. Starting with DR4 of the Legacy Surveys,  PS1 positions of these sources were replaced with the Gaia DR1 catalog positions; i.e.\ post-DR3 astrometry is tied to Gaia.  
The astrometric residuals for bright stars relative to their Gaia DR1 catalog positions are shown in Figure~\ref{fig:astrometryMzLS}, \ref{fig:astrometryBASS} and \ref{fig:astrometryDECaLS}. MzLS and DECaLS have  rms scatters of $\approx20$\,mas, with BASS showing slightly larger residuals. 
The residual scatter, outliers, and asymmetries visible in the distributions shown in Figures~\ref{fig:astrometryMzLS}, \ref{fig:astrometryBASS} and \ref{fig:astrometryDECaLS} are likely due to the following reasons: the higher order and small-scale pixel-level distortions, which are larger in the thick, deep depletion CCDs in the Mosaic-3 and DECam cameras; (2) the lack of proper motions in the Gaia DR1 catalog, which exaggerates the scatter and causes outliers because of the difference in epoch between the two images; (3) the plots shown include all data, irrespective of observed image quality. 
Corrections for the offsets due to the higher order pixel-level distortions and modelling of stars with known proper motions will be incorporated into \tractor\ modeling in future data releases.

\subsection{Photometric Calibration}

The role of the CPs in photometric calibration is to remove any spatial variation in the photometric response of each CCD (i.e., to ``flatten’’ each CCD) and to then estimate the conversion factor from analog digital units recorded by each CCD to photoelectrons.  
The CPs also provide the data quality masks and weight maps which are used for in the subsequent source detection and photometric calibration steps.  

The Legacy Surveys are designed so that each part of the footprint is observed in photometric conditions at least once, and most of the footprint is observed in photometric conditions two or more times.  Efforts to observe at the lowest possible airmasses and avoid the Moon drive an observing plan that features a rich set of overlaps between observations on different nights.  Comparison of observations of the same stars on different nights and at different airmasses then enables determination of the system throughput and the transparency of the atmosphere for each photometric night of the survey.  This procedure is the basis of the photometric calibration of the SDSS \citep{padmanabhan2008}, as well as subsequent surveys like PS1 \citep{schlafly2012} and DES \citep{burke2018}.  Observations on non-photometric nights will be calibrated by matching directly to overlapping observations taken on photometric nights.

The current photometric calibration for the Legacy Surveys (for all data releases through DR6) is, however, tied to the PS1 DR1 photometry through a set of color transformation equations. The magnitudes of PS1 DR1 catalog sources are first converted to the ``native'' system for each telescope+camera+filter, and the transformations are as follows:
\begin{eqnarray}
g_{\rm DECaLS} = g_{\rm PS1} + 0.00062 + 0.03604(g-i)_{\rm PS1} + 0.01028(g-i)_{\rm PS1}^2 - 0.00613(g-i)_{\rm PS1}^3 \\
r_{\rm DECaLS} = r_{\rm PS1} + 0.00495 - 0.08435(g-i)_{\rm PS1} + 0.03222(g-i)_{\rm PS1}^2 - 0.01140(g-i)_{\rm PS1}^3 \\
z_{\rm DECaLS} = z_{\rm PS1} + 0.02583 - 0.07690(g-i)_{\rm PS1} + 0.02824(g-i)_{\rm PS1}^2 - 0.00898(g-i)_{\rm PS1}^3 \\
g_{\rm BASS} = g_{\rm PS1} + 0.00464 +0.08672(g-i)_{\rm PS1}  -0.00668(g-i)_{\rm PS1}^2  -0.00255(g-i)_{\rm PS1}^3 \\
r_{\rm BASS} = r_{\rm PS1} +0.00110 -0.06875(g-i)_{\rm PS1} +0.02480(g-i)_{\rm PS1}^2 -0.00855(g-i)_{\rm PS1}^3 \\
z_{\rm MzLS} = z_{\rm PS1} +0.03664 -0.11084(g-i)_{\rm PS1} +0.04477(g-i)_{\rm PS1}^2 -0.01223(g-i)_{\rm PS1}^3
\label{eqn:PS1LScolors}
\end{eqnarray}
These color transformations are measured empirically by comparing the Legacy Surveys and PS1 catalog data for stars with magnitudes $<$21.5 AB mag and colors $0<(g-i)_{\rm PS1}<2.9$, and the absolute calibration is determined using the CALSPEC database\footnote{Specifically, the November 2017 version of the release. See \url{http://www.stsci.edu/hst/observatory/crds/calspec.html} for details.} \citep[see][and references therein]{calspec2017}. The DECaLS transformations used above are the same as those determined by \citet{Schlafly2018} for the DECaPS Galactic Plane Survey. The BASS and MzLS transformations were determined in a similar manner, using unresolved sources selected from PS1 DR1 with well-measured photometry (i.e., no flags) with colors in the range $0 < (g-i)_{\rm PS1}<2.9$.   Constant terms in the calibration are intended to place the Legacy Surveys on the AB magnitude system \citep{Oke83}, and were derived from comparison of the empirical transformations with synthetic transformations of calibrated Hubble Space Telescope standard stars, given the system throughputs of the DECam, BASS, and MzLS surveys. (The photometric transformations between the SDSS $grz$ magnitudes and the Legacy Surveys' magnitudes are presented in Appendix~\ref{sec:sdssvsls}.)

We estimate a zero point for each CCD independently, by (1) detecting sources on the pipeline-reduced data; measuring their instrumental magnitudes; (2) matching to the subset of PS1 DR1 catalog sources selected as calibrators; and then (3) comparing the instrumental magnitudes to the color-transformed PS1 DR1 magnitudes (i.e., as per Equations~1--\ref{eqn:PS1LScolors}). This procedure results in zero points for each CCD tied to the global PS1 calibration, but corrected to the ``native'' photometric frame for each individual survey.

In the future, we will migrate to an internal photometric calibration that will rely solely on  data from  each of the Legacy Surveys. 
The large network of repeat observations on different photometric nights enables the construction of a detailed description of the throughput of the various imaging systems used in the Legacy Surveys.  We plan to not only measure overall system zero points and (grey) atmospheric transparency in each band on each night \citep[cf.,][]{padmanabhan2008}, but also to determine how sensitivity varies within and among the different CCDs of each system as a function of time, \citep[cf.][]{schlafly2012, Schlafly2018}.  We can also determine and ameliorate systematic problems with aperture correction.  Well-calibrated optical colors are now available from PS1 and Gaia, which will make it straightforward to remove systematic chromatic errors stemming from the different colors of stars and the varying effective throughput of the imaging system in different conditions \citep{li2016,burke2018}.  
The DECam Plane Survey, which used a similar three-pass strategy to DECaLS, obtained 6-8 mmag precision for bright stars \citep{Schlafly2018}, without accounting for color-dependent calibration terms; we anticipate similar photometric precision for the Legacy Surveys data. 

\section{Inference Modeling with \emph{The Tractor}}
\label{sec:tractor}

All the source catalogs from the Legacy Surveys project are constructed using \tractor.
Co-author D.~Lang has developed \tractor\footnote{Publicly available at \url{https://github.com/dstndstn/tractor}}
as a forward-modeling approach to perform source extraction on pixel-level data.
This algorithm is a statistically rigorous approach to fitting the
differing PSF and pixel sampling of the different imaging data that comprise the Legacy Surveys. This approach is particularly useful given the wide range in PSF shape and size exhibited by the Legacy Surveys data: the optical
data have a typical PSF of $\approx 1$ arcsec; and the {\it WISE} PSF FWHM is
$\approx 6$\,arcsec in W1--W3 and $\approx 12$\,arcsec in W4.

For the Legacy Surveys, we have created a post-processing catalog generation pipeline called {\it legacypipe}\footnote{Publicly available at \url{https://github.com/legacysurvey/legacypipe}}, which wraps \tractor, and which proceeds as follows. The Legacy Surveys footprint is analyzed in $0.25^\circ\times0.25^\circ$ regions called ``bricks’’. We first identify all the CCDs that overlap a given brick, and each CCD is analyzed to estimate and subtract the sky. The initial sky estimate is computed by first subtracting a per-CCD median value of the unmasked pixels, then estimating a sliding median every 512 pixels on a box size of 1024 pixels, and fitting the result with a two-dimensional spline. This initial sky is biased by sources, but does remove slow variations in the sky background. Subtracting this initial sky model, we compute a 5-pixel boxcar-smoothed image, detect and mask pixels above 3 sigma (in boxcar-smoothed sigmas) plus a 3-pixel margin, and recompute the spline background estimate using the remaining unmasked pixels. This iteration results in a sky estimate less biased by sources in the image. 
The PSF for each CCD is then estimated on the sky-subtracted image using PSFEx \citep{psfex}, and each individual sky-subtracted CCD is convolved with its own PSF in order to facilitate source detection. We then create five separate stacks for the purpose of source detection: a weighted sum of all the (PSF-convolved) CCDs in a given band (resulting in three such stacks); a weighted sum of all three bands to optimize for a ``flat'' SED (i.e., zero AB mag color); and a weighted sum of all three bands to optimize for a ``red'' SED (i.e., with colors $g-r=1$~mag and $r-z=1$~mag). While these image stacks are weighted sums of the convolved images, the input images are not all convolved to a common PSF. Next, we detect sources on the three individual band image stacks and the two $grz$ image stacks using a simple thresholding algorithm, selecting sources above 6$\sigma$. This process identifies almost all sources in the images to faint magnitudes. The details of the entire legacypipe pipeline (and {\it The Tractor}) will be presented in a forthcoming paper (Lang et al., in prep).

Next, we detect sources on the individual-band image stacks and the two $grz$ image stacks using a simple thresholding algorithm, selecting sources above 6$\sigma$. This process identifies almost all sources in the images to faint magnitudes.
ach source is then modeled by \tractor, which takes as input the NOAO pipeline-reduced individual images from
multiple exposures in multiple bands, with different seeing in each.
For each astronomical source, a source model is fit simultaneously to the pixel-level data 
of all images containing the source.
\tractor\ models each source using a small set of parametric light profiles: a delta function (for point sources); a deVaucouleurs $r^{-1/4}$ law; an exponential disk; or a ``composite'' deVaucouleurs plus exponential. The best fit model is determined by convolving each model with the specific PSF for each individual exposure, fitting to each image, and minimizing the residuals for all images. The PSF for each optical image is constructed using PSFEx \citep{psfex}. We make the assumption that the model is the same across all the bands. 
Thus, if a source is determined
to be a point source, it is modeled as a point source in every band
and every exposure and its catalog photometry is based on this model.
Alternatively, if the source is spatially extended, then the same
light profile (an exponential disk, de Vaucouleurs, or combination) is consistently fit to all images in order to determine the best-fit source position, source shape parameters and photometry\footnote{Further details regarding the catalog construction and source parameter extraction can be found at the Legacy Surveys' \href{http://legacysurvey.org/dr6/description/}{website describing the latest data release}.}.

\tractor\ model fits are determined using only the optical $grz$ data. The mid-infrared photometry for each optically-detected source is then determined by forcing the location and shape of the model, convolving with the {\it WISE} PSF and fitting to the {\it WISE} stacked image. This ``forced photometry'' approach allows us to deblend any confused {\it WISE} sources by using the higher-spatial-resolution optical data, but also limits the Legacy Surveys catalogs to only contain {\it WISE} photometry for sources that are detected at optical wavelengths. 
The procedure described produces object positions, fluxes
and colors that are consistently measured across the three Legacy Surveys. 

\begin{figure}[!th]
\centering
\includegraphics[width=6.5in]{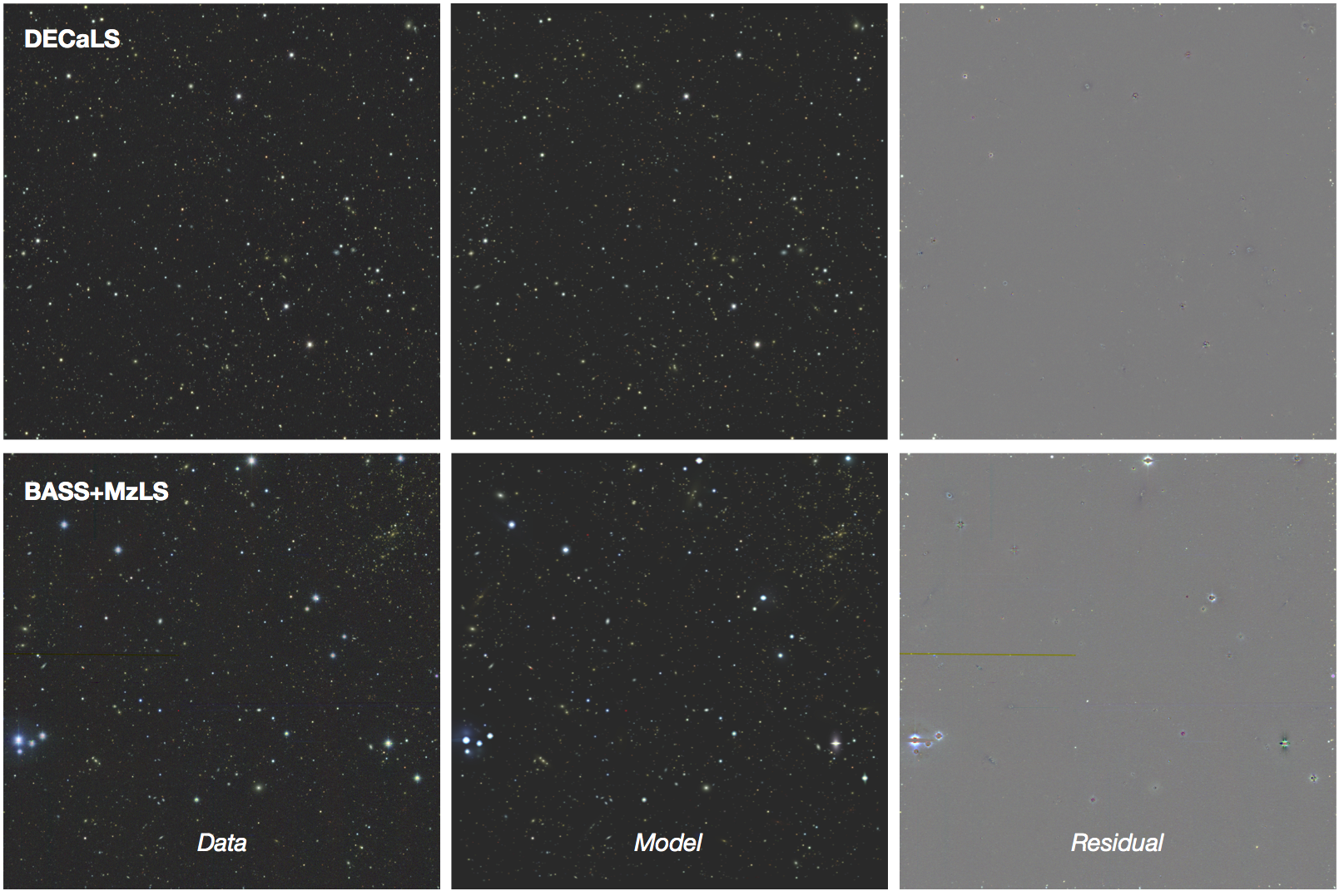}
\caption{
Example ``bricks" covering $0.25\times0.25$~deg$^2$ from the DECaLS survey (top row; brick 2212p085) and the MzLS and BASS surveys (bottom row; brick 1689p532). From left to right, the panels show the actual $grz$ imaging data, the rendered model based on \tractor\ catalog of the region, and the residual map. \tractor\ catalog represents an inference-based model of the sky that best fits the observed data. Readers can explore the data, models and residual images in more detail using the Legacy Surveys {\it Imagine} sky viewer at \url{http://legacysurvey.org/viewer}
\label{fig:DECaLStrac}}
\end{figure}

Figure~\ref{fig:DECaLStrac} shows examples of how \tractor\ is being applied to the  Legacy Surveys. The footprint of the Legacy Surveys  is divided into ``bricks'' of size $0.25^\circ\times0.25^\circ$, and a model of the sky within each brick is computed using all CCDs that contribute data within that brick. The three sets of vertical panels show: the $grz$ image data for a brick; the rendered \tractor\ model; and the residual image (i.e., ${\rm data} - {\rm model}$). While most of the faint sources are well fit by the parametric models we used for the Legacy Surveys, more significant residuals are seen associated with very extended galaxies and the halos of bright, saturated stars. 

\tractor\ and our source detection algorithms do result in the catalog containing a small fraction of spurious sources. These are primarily due to inadequately masked particle events or \href{http://legacysurvey.org/viewer?ra=162.3716&dec=55.9808&zoom=14&layer=mzls+bass-dr6&sources-dr6}{satellite trails}, single-exposure detections of transient sources (primarily \href{http://legacysurvey.org/viewer?ra=242.3810&dec=8.6956&zoom=15&layer=decals-dr5}{asteroids}), and sources identified in the extended \href{http://legacysurvey.org/viewer?ra=163.7356&dec=55.8671&zoom=15&layer=mzls+bass-dr6}{scattered light halos or diffraction spikes associated with bright stars}, or in the \href{http://legacysurvey.org/viewer?ra=161.3316&dec=55.9613&zoom=15&layer=mzls+bass-dr6}{diffuse emission associated with large galaxies}. In addition, spatially large, extended sources with complex morphologies (e.g., \href{http://legacysurvey.org/viewer?ra=219.3117&dec=38.4544&zoom=13&layer=mzls+bass-dr6}{large galaxies}) and crowded fields (e.g., \href{http://legacysurvey.org/viewer?ra=229.6407&dec=2.0808&zoom=13&layer=decals-dr5}{globular clusters} and \href{http://legacysurvey.org/viewer?ra=132.8378&dec=11.8321&zoom=12&layer=decals-dr5}{open star clusters}) are poorly modeled by \tractor. Finally, a small number of sources are missed by \tractor\ catalog; these are primarily \href{http://legacysurvey.org/viewer?ra=325.6999&dec=1.0124&zoom=15&layer=decals-dr5}{very low surface brightness diffuse sources} or \href{http://legacysurvey.org/viewer?ra=243.4308&dec=7.1181&zoom=16&layer=decals-dr5-resid}{sources lying close to a (typically brighter) star or galaxy}. 

\begin{figure}[!ht]
\centering
\includegraphics[width=0.49\textwidth]{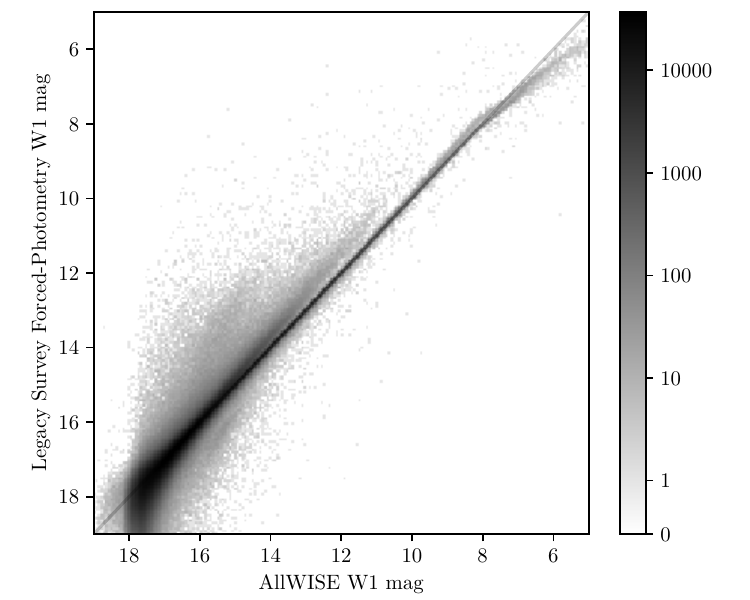}
\includegraphics[width=0.49\textwidth]{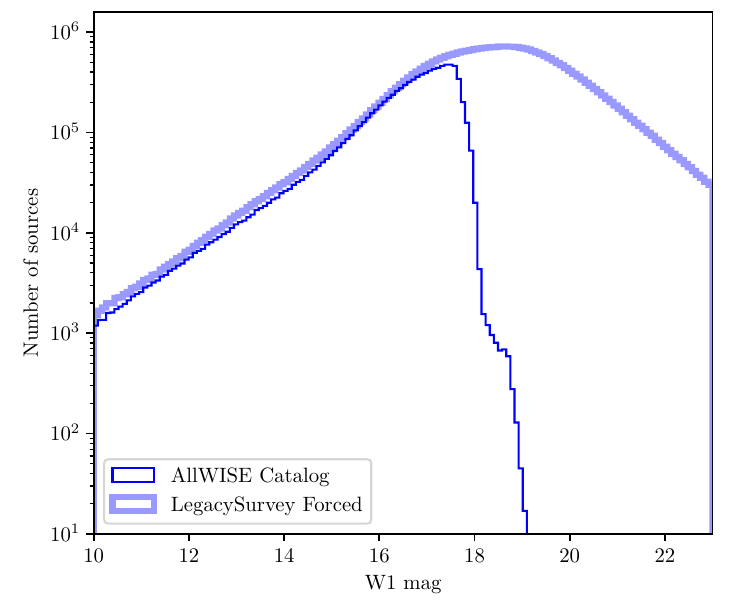}
\caption{Forced photometry with \tractor\ code, using
  information from Legacy Surveys detections and light profiles, allows us to measure the
  mid-infrared flux from objects in the {\it WISE} images to below the {\it
    WISE} detection limit.  {\it Left panel:} A comparison of the $W1$ photometric measurements in the Legacy Surveys' catalog (derived using \tractor) with those in the ALLWISE catalog; the greyscale shows the relative density of points. The photometry agrees well for mid-infrared bright
  objects that are detected in the AllWISE catalog.  The widening
  locus below $W1\sim14$ is due to \tractor\ photometry treating larger
  objects as truly extended, in contrast to the point-source-only
  assumptions in the public AllWISE catalog.  {\it Right panel:} The number counts in the Legacy Surveys' catalog compared with those from AllWISE, demonstrating the increased depth made possible from using
  \tractor.  By using optical imaging from Legacy Surveys to detect
  objects, photometry is measured for objects that are well
  below the detection limit of the AllWISE catalog.}
  \label{fig:tractor-bright}
\end{figure}

\begin{figure}[!th]
\centering
\includegraphics[width=0.49\textwidth]{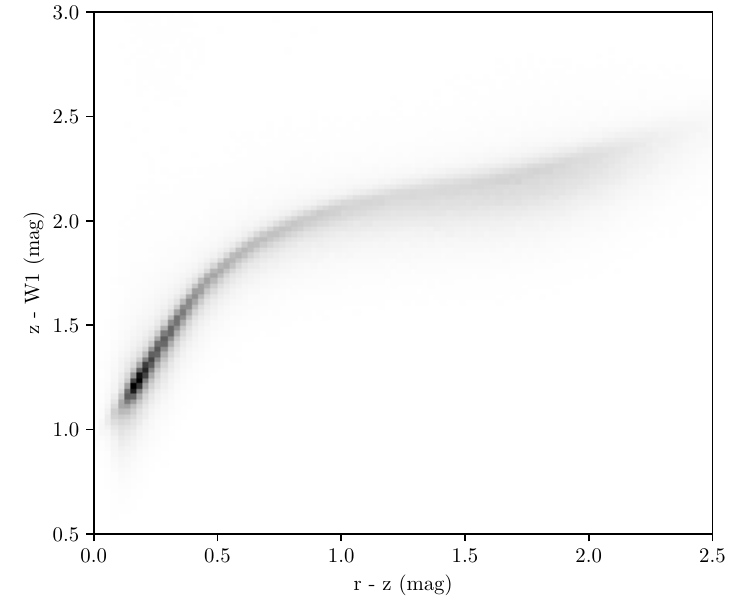}
\includegraphics[width=0.49\textwidth]{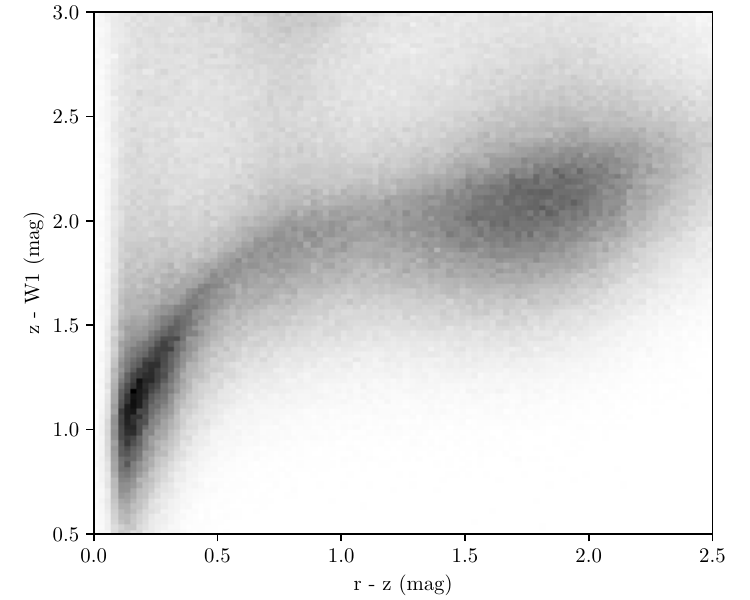}
\caption{Forced photometry using \tractor, contrasted with traditional
  ``catalog-matching.''  {\it Left:} Color-color diagram of
  catalog-matched PSF sources from the Legacy Surveys and AllWISE
  catalogs.  {\it Right:} Color-color diagram of PSF sources from the
  Legacy Survey catalogs that have \emph{no} matching object in the
  AllWISE catalog.  Many of these will be sources that are well
  detected in the Legacy Surveys optical imaging, but below the
  detection threshold in the {\it WISE} imaging.  We have applied a cut to
  show only {\it WISE} flux measurements above $3 \sigma$; with this cut, we
  nearly double the number of sources with measured {\it WISE} fluxes.
  The distributions shown in both panels are similar, demonstrating that the
  forced photometry measurements of the faint mid-infrared sources make astrophysical 
  sense, despite being noisier.}
  \label{fig:tractor-cmd}
\end{figure}

The ``forced photometry" approach we use to measure the mid-infrared photometry from the {\it WISE} images allows us to detect fainter sources than a traditional approach while preserving the photometric reliability. For bright objects that were cleanly detected by {\it WISE} alone (and recorded in the AllWISE catalog), the pixel-level measurements are consistent with catalog-level measurements
(see Figure \ref{fig:tractor-bright}, left panel).  However, we are also able to
measure the fluxes of significantly fainter objects, as well as 
study collections of objects that are blended in the {\it WISE}
images but that are resolved in the optical images. The increased level of source counts in the force-photometered data observed at bright {\it WISE} magnitudes (i.e., $W1\lesssim 16$~mag) in the right panel of figure~\ref{fig:tractor-bright} is due to sources detected in the vicinity of bright stars or galaxies; $\approx$50\% of these sources are real objects, and the remaining are spurious detections due to the halos or diffraction spikes around bright stars or the poorly-modeled extended light of galaxies.  All are located near other brighter targets, which is why they are compromised.
We are currently working on improving the models for future data releases. 
Figure \ref{fig:tractor-cmd} compares a traditional 
optical-infrared color-color diagram, based on matching sources between catalogs at different wavelengths, to the photometry derived from our {\it WISE} forced photometry, which requires no such matching. This demonstrates how \tractor\ greatly increases the access to mid-infrared photometry for targets fainter than the AllWISE catalog detection limits, albeit with increased scatter. 
We have verified the reliability of the forced photometry detections and measurements by comparing \tractor\ catalog results with those from deep {\it Spitzer} data in the COSMOS field \citep[i.e., the S-COSMOS catalog from][]{scosmos}. Defining reliability as the fraction of {\it Spitzer} sources recovered, we find that the reliability is $\ge$95\% for sources with W1 or W2 signal-to-noise ratios of $\ge5$, corresponding to 21.3 and 20.4 AB mag respectively. We have also compared the photometric measurements in our catalogs with those from S-COSMOS and find that they are in good agreement for the $\ge 5\sigma$ detections.


\vspace{0.2in}

\tractor\ should improve target selection for all
DESI target classes by allowing information from low signal-to-noise ratio measurements to be utilized. \tractor\ is particularly important for targeting QSOs.
Up to 15\% of QSO spectra exhibit broad absorption lines that
potentially reduce the measured flux in broad-band imaging. In addition, high-redshift QSOs at $z\ge5.0$ ($\ge6.9$) will drop out of the $g$-band ($g$- and $r$-bands) completely \citep[e.g., see][]{banados2018}. 
Finally, the $5\sigma$ optical limit at the extremes of DESI targeting corresponds to
a $< 5\sigma$ limit in {\it WISE} for QSOs.
\tractor\ successfully differentiates between the QSOs that
are detected in {\it WISE} and those that are not detected, whereas
traditional ``catalog-matching'' approaches would not be successful.

The SDSS-IV/eBOSS \citep{Daw16}, which began observations in July 2014, also utilized \tractor\ and 
the {\it WISE} component of the Legacy Surveys to target LRGs \citep{Pra16} and QSOs \citep{Mye15}.
For these eBOSS targets, \tractor\ provided forced photometry based
upon galaxy profiles measured by the SDSS imaging pipeline. Those profiles were
convolved with the {\it WISE} PSF, and then a linear fit
was performed on the full set of {\it WISE} imaging data.
The result was a set of
mid-infrared flux estimates for all SDSS objects, constructed so that the sum of
flux-weighted profiles best matched the {\it WISE} images \citep{Lang2014,Lang16}. The Legacy Surveys use 
the same fitting approach, using the deeper DECaLS, MzLS and BASS optical images in place of the SDSS images. 

\tractor\ catalogs 
include source positions, fluxes, shape parameters, and morphological quantities that can be used to 
discriminate extended sources from point-sources, together with errors on these quantities.  \tractor\ catalogs are vetted
for DESI target selection using a series of image validation tests (as in e.g., Appendix \ref{sec:hscvsls}).

\section{Data Releases from the Legacy Surveys\label{sec:status}}

The Legacy Surveys are being run as completely public projects. All raw optical imaging data are made available as soon as they are transferred from each telescope to the NOAO Science Archive\footnote{\url{http://archive.noao.edu},\url{https://datalab.noao.edu/decals/ls.php}}. The data transfer occurs within minutes for DECam and Mosaic-3 data, and by the following morning for 90Prime data. Pipeline processed data are made public as soon as the reductions are completed, typically within a week of the observations. Finally, catalogs based on \tractor\ and cross-matched data are released twice a year. All data (images, coadds, catalogs and supplementary material) are available at the NOAO Science Archive and through the Legacy Surveys portal hosted at NERSC\footnote{\url{http://legacysurvey.org}}. All the code used for creating the catalogs is publicly 
available\footnote{\url{https://github.com/legacysurvey/legacypipe} and \url{https://github.com/dstndstn/tractor}}. 
In addition, the BASS data are independently processed and released by the BASS team \citep[e.g., see][for details]{bassDR1,bassDR2}. 

Tables \ref{tab:observing} and \ref{tab:drs} show the observing schedule at the Blanco, Bok, and Mayall telescopes and our data release milestones. The data releases provide required deliverables such as object catalogs, depth maps, and co-added images, models, and residuals (see Table~\ref{tab:dataprods}; see \url{http://legacysurvey.org} for details regarding the data release contents). 

As of December 2018, the two most recent data releases are 
DR6 (containing all the MzLS and BASS data obtained through 2017 December 9 and 2017 June 25, respectively) 
and 
DR7 (containing all the DECaLS data obtained through 2018 March); 
see Table~\ref{tab:drs}. Together, DR6+DR7 jointly cover nearly the entire DESI footprint and have significant overlap. The DR7 data release presents DECaLS data over $\approx$9,766~deg$^2$ in $g$-band, 9,853~deg$^2$ in $r$-band, and 10,610~deg$^2$ in $z$-band, with 9,298~deg$^2$ of coverage in all three bands. There are approximately 835 million unique sources in DR7 spread over 180,102 bricks. The DR6 data release presents MzLS+BASS data covering $\approx$4,400~deg$^2$ in $g$-band, 4,400~deg$^2$ in $r$-band, and 5,300~deg$^2$ in $z$-band, with $\approx$3,900~deg$^2$ with three-band coverage. There are approximately 310 million unique sources in DR6 spread over 92,287 bricks. Approximately 7500 bricks overlap between DR6 and DR7, each containing more than 1000 objects.

\begin{figure}[!th]
\includegraphics[width=1.1\textwidth]{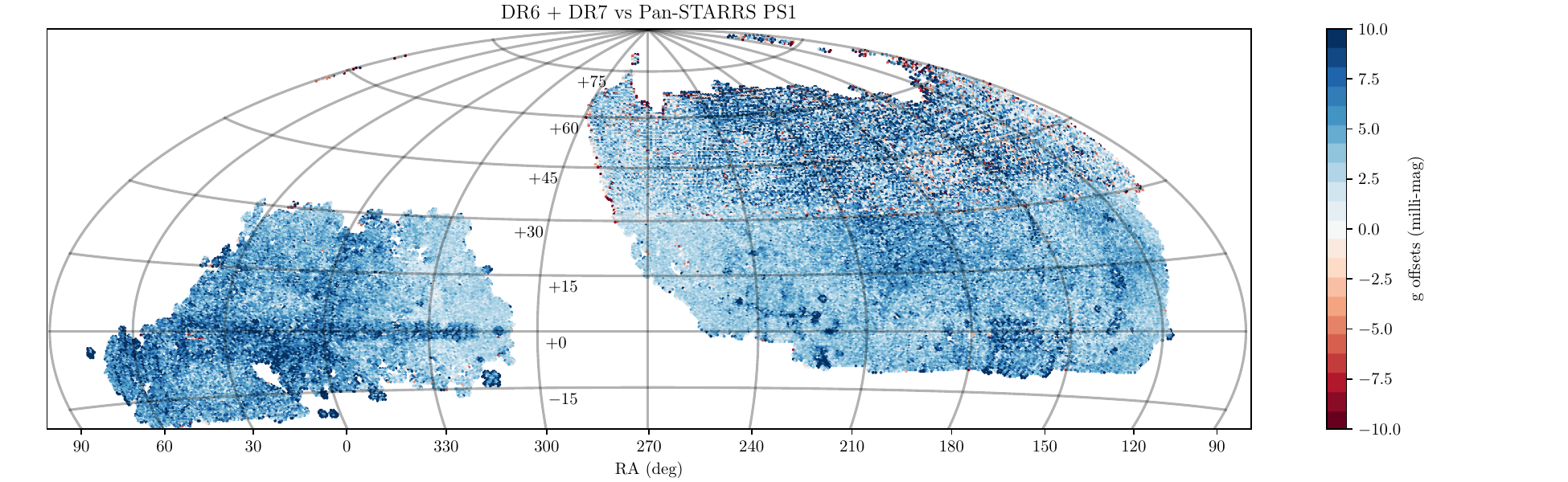}
\includegraphics[width=1.1\textwidth]{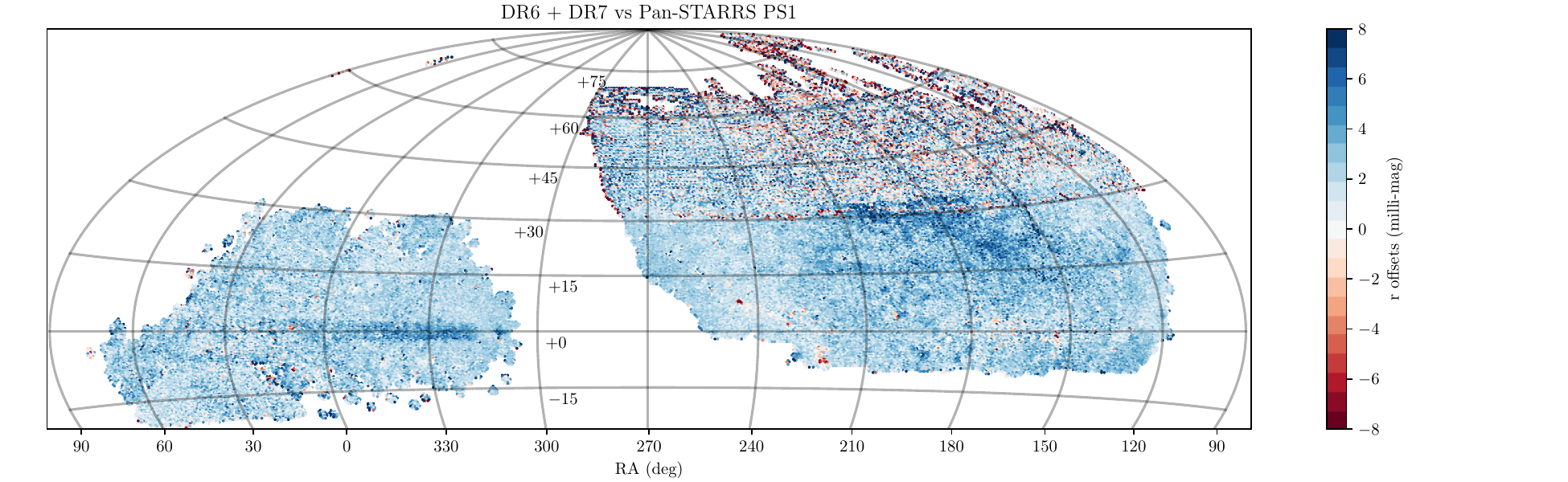}
\includegraphics[width=1.1\textwidth]{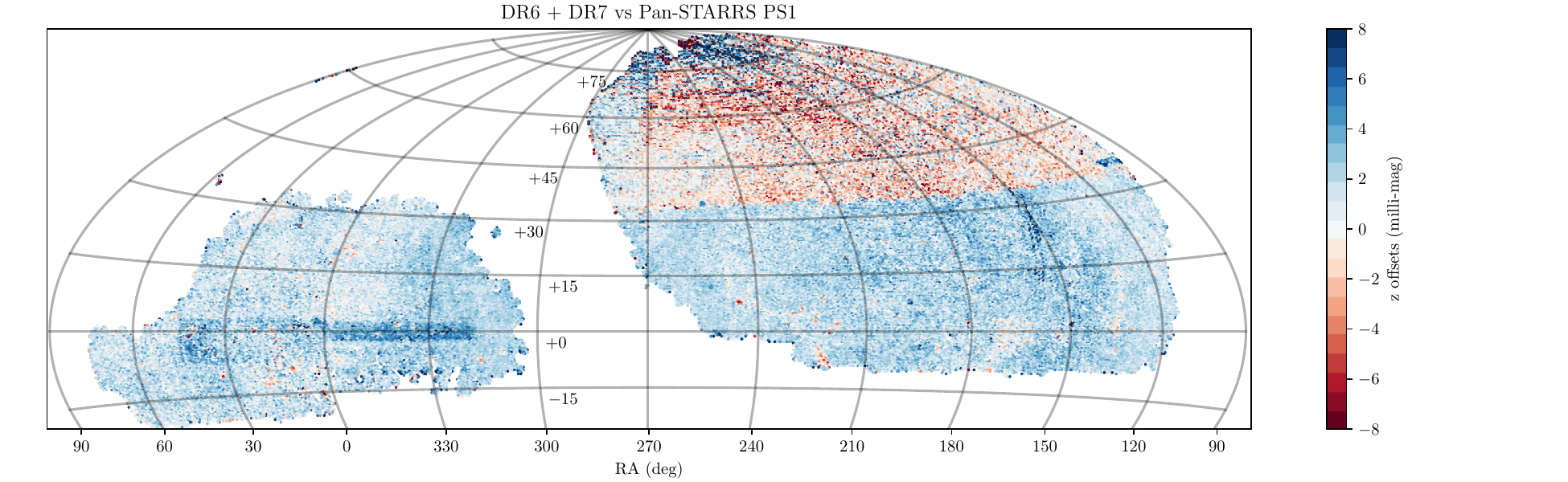}
\caption{The spatial distribution (at 0.5$^\circ$ resolution) of photometric residuals in the $g$- (top), $r$- (middle), and $z$-band (bottom) in the Legacy Surveys DR6 and DR7 releases, computed relative to the PanSTARRS survey. The comparison presented here is between stellar (i.e., unresolved) objects in the Legacy Surveys and PS1 catalogs, after correcting for the difference in the color terms between the two surveys. 
\label{fig:skyzpt}}
\end{figure}

\begin{figure}[!th]
\includegraphics[width=1.1\textwidth]{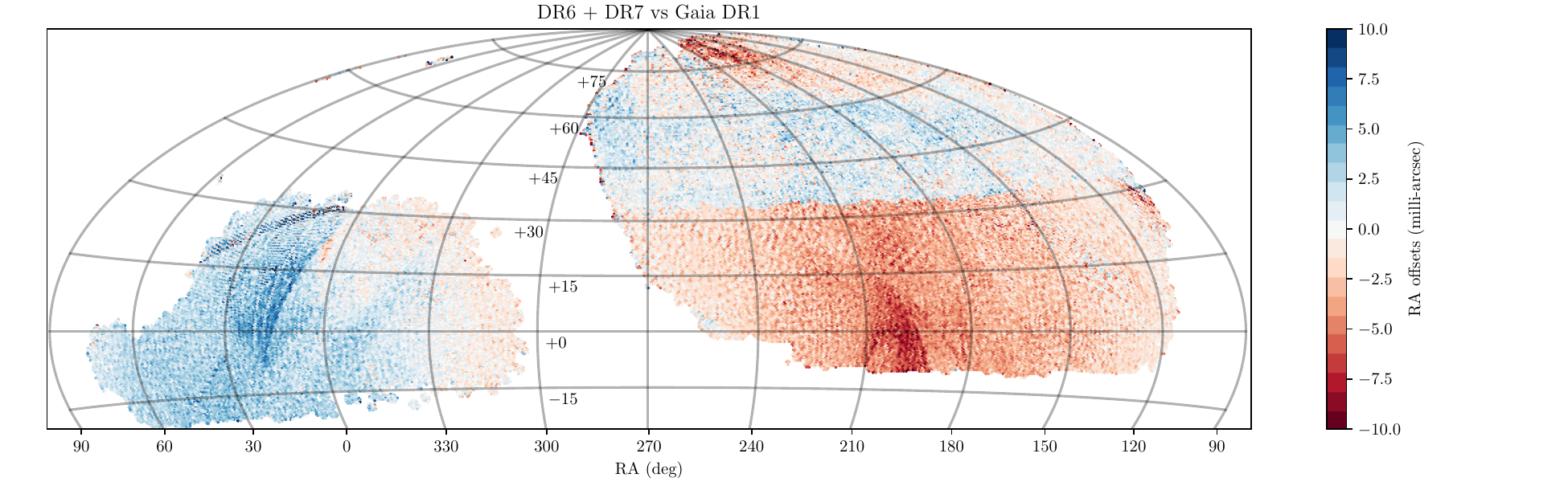}
\includegraphics[width=1.1\textwidth]{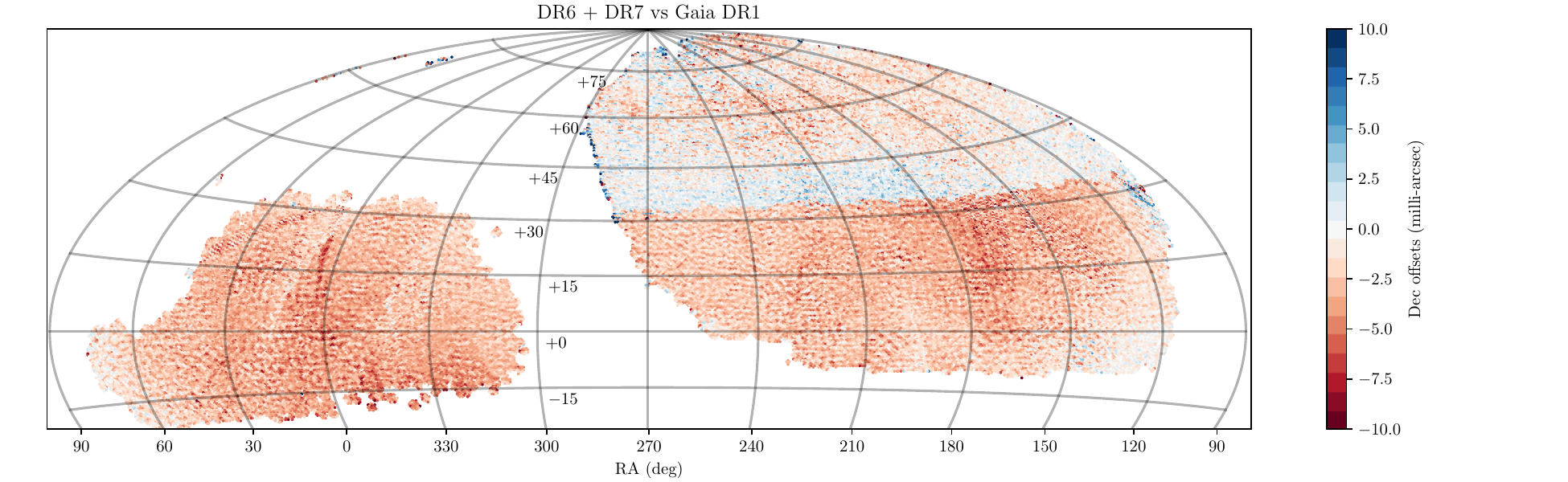}
\includegraphics[width=1.1\textwidth]{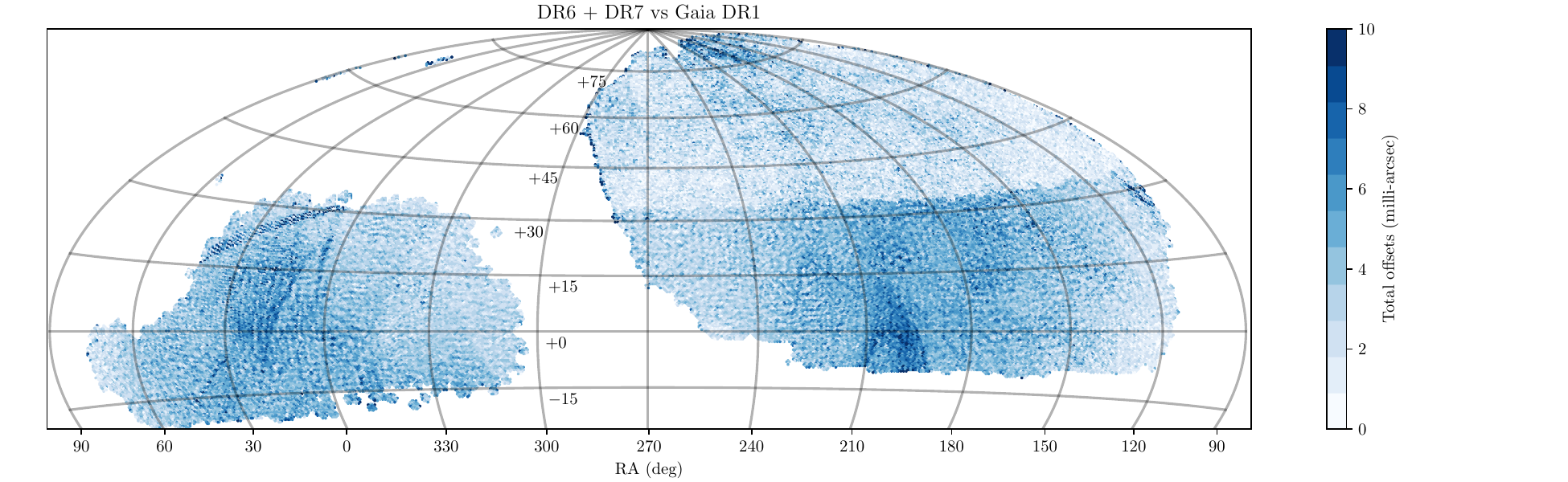}
\caption{The spatial distribution (at 0.5$^\circ$ resolution) of astrometric residuals in the Legacy Surveys DR6 and DR7 releases, computed relative to Gaia DR1 survey. While the rms scatter in the residuals is small ($\sigma_{\rm RA,Dec}\approx2$~mas), there remain spatially coherent systematic offsets at the $<\pm5$~mas level in the different survey regions. The systematic offset at the Declination +34$^\circ$ boundary are due to differences in the way the astrometric zero points were computed in the DR6 and DR7 releases, and will be corrected in the Legacy Surveys DR8 release.
\label{fig:skyastrom}}
\end{figure}

\begin{figure}[!th]
\includegraphics[width=1.1\textwidth]{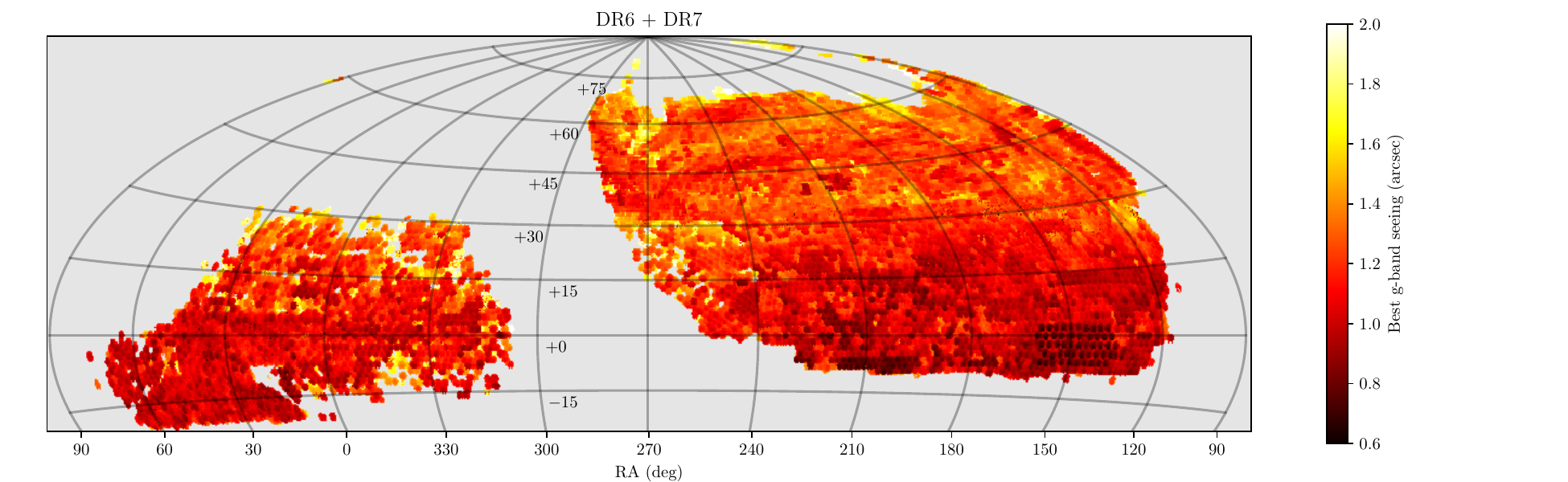}
\includegraphics[width=1.1\textwidth]{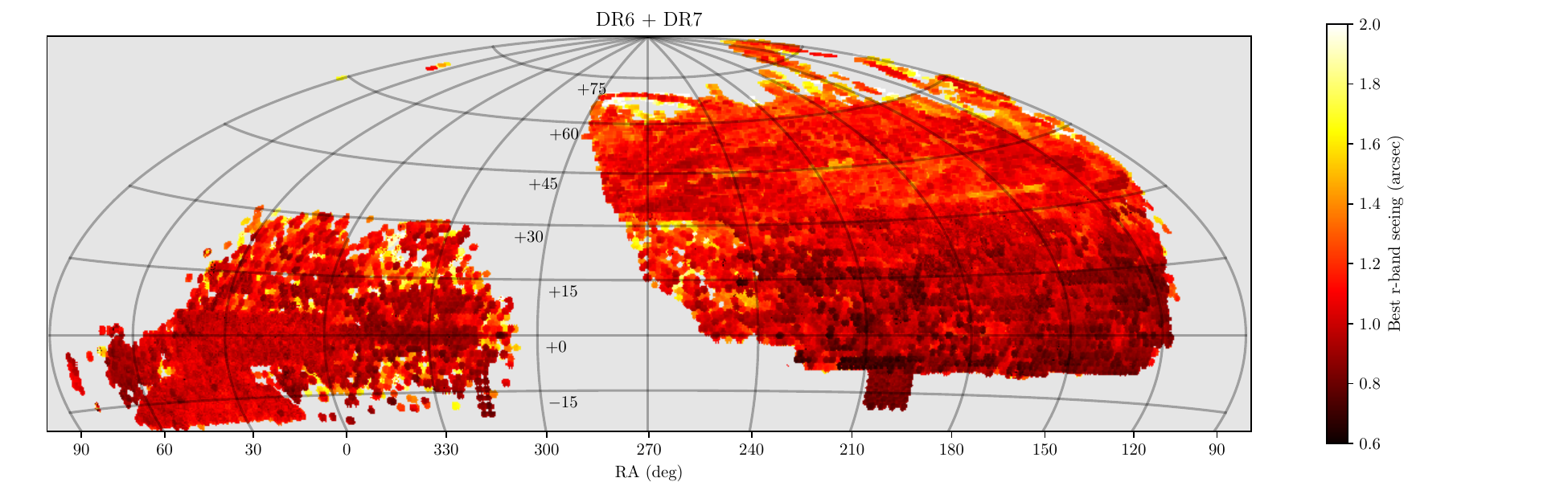}
\includegraphics[width=1.1\textwidth]{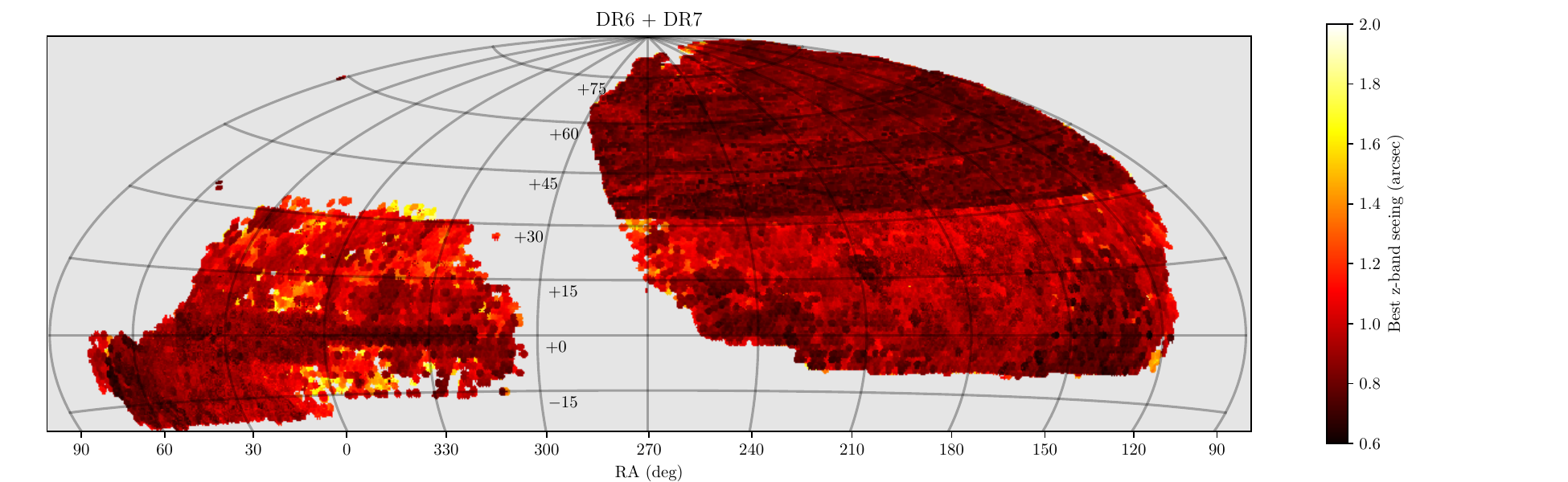}
\caption{The spatial distribution (at 0.5$^\circ$ resolution) of image quality represented by the best-seeing data in the $g$- (top), $r$- (middle), and $z$-band (bottom) in the Legacy Surveys DR6 and DR7 releases. The \tractor\ source modeling is dependent on the data with the best delivered image quality at any given location. The DECaLS survey (covering the region south of $\delta\sim+34^\circ$ is still incomplete, as reflected by the variable image quality in much of this region. The DR6 release of MzLS also did not include the full MzLS dataset. 
\label{fig:zpsf}}
\end{figure}

\begin{figure}[!th]
\includegraphics[width=0.32\textwidth]{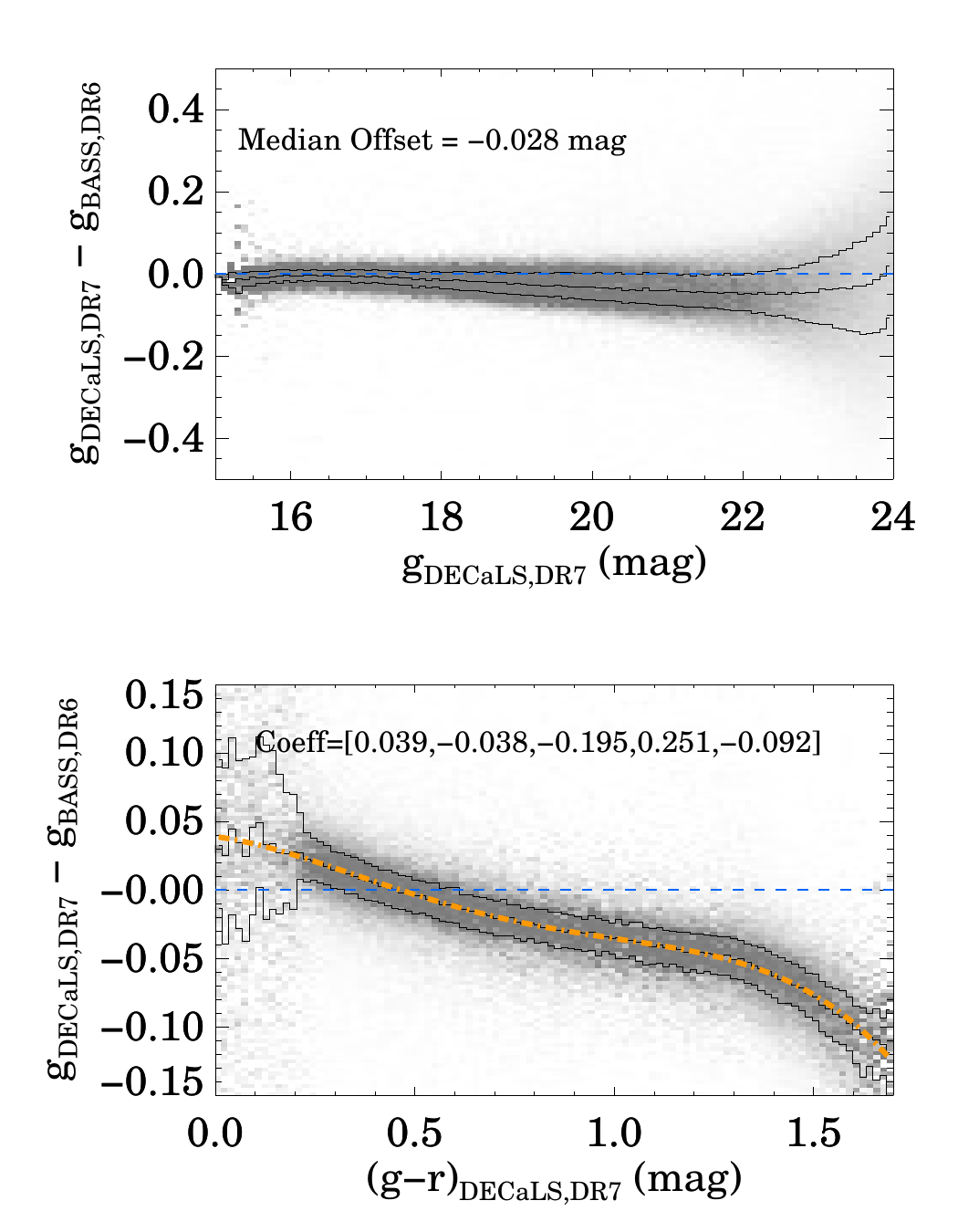}
\includegraphics[width=0.32\textwidth]{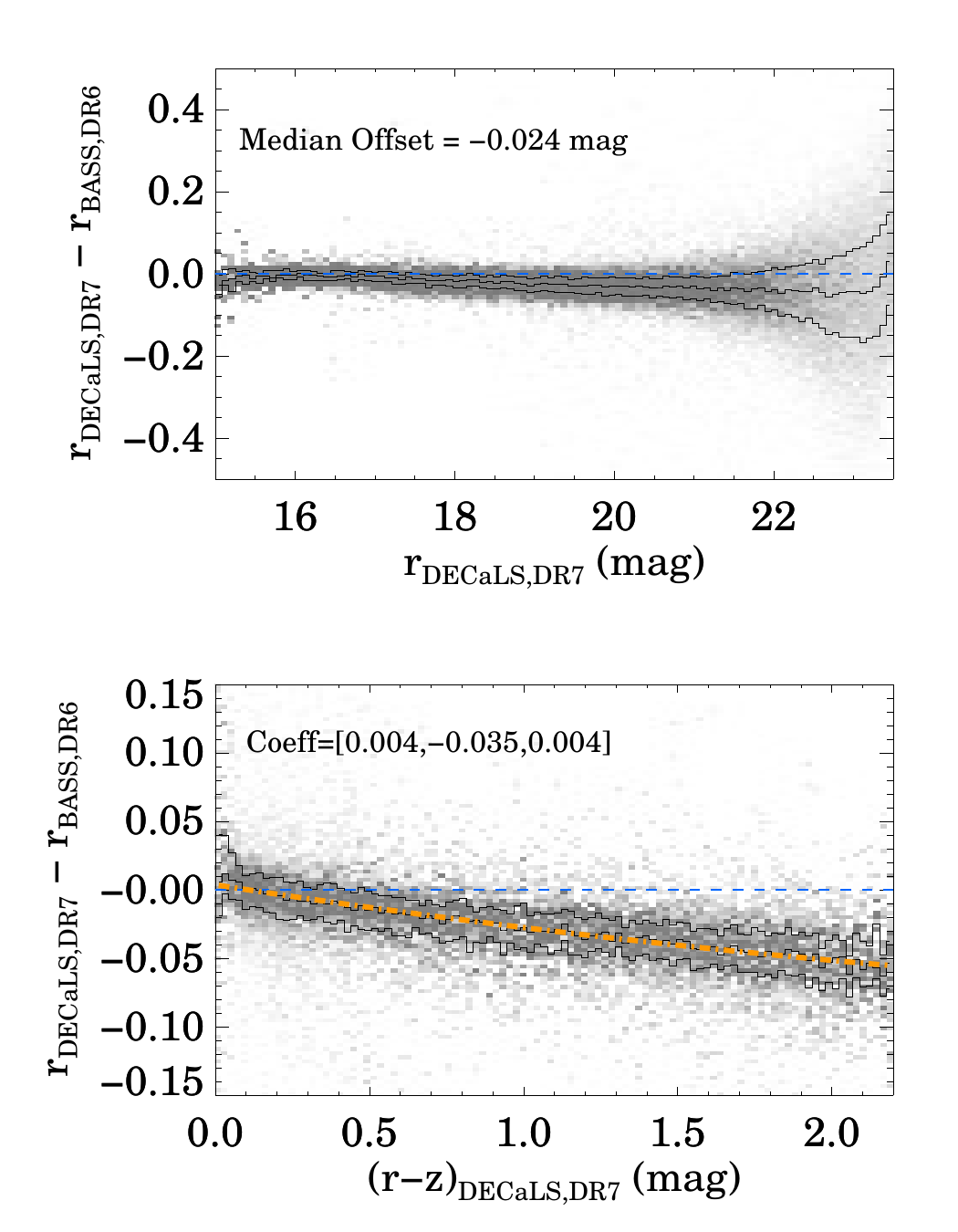}
\includegraphics[width=0.32\textwidth]{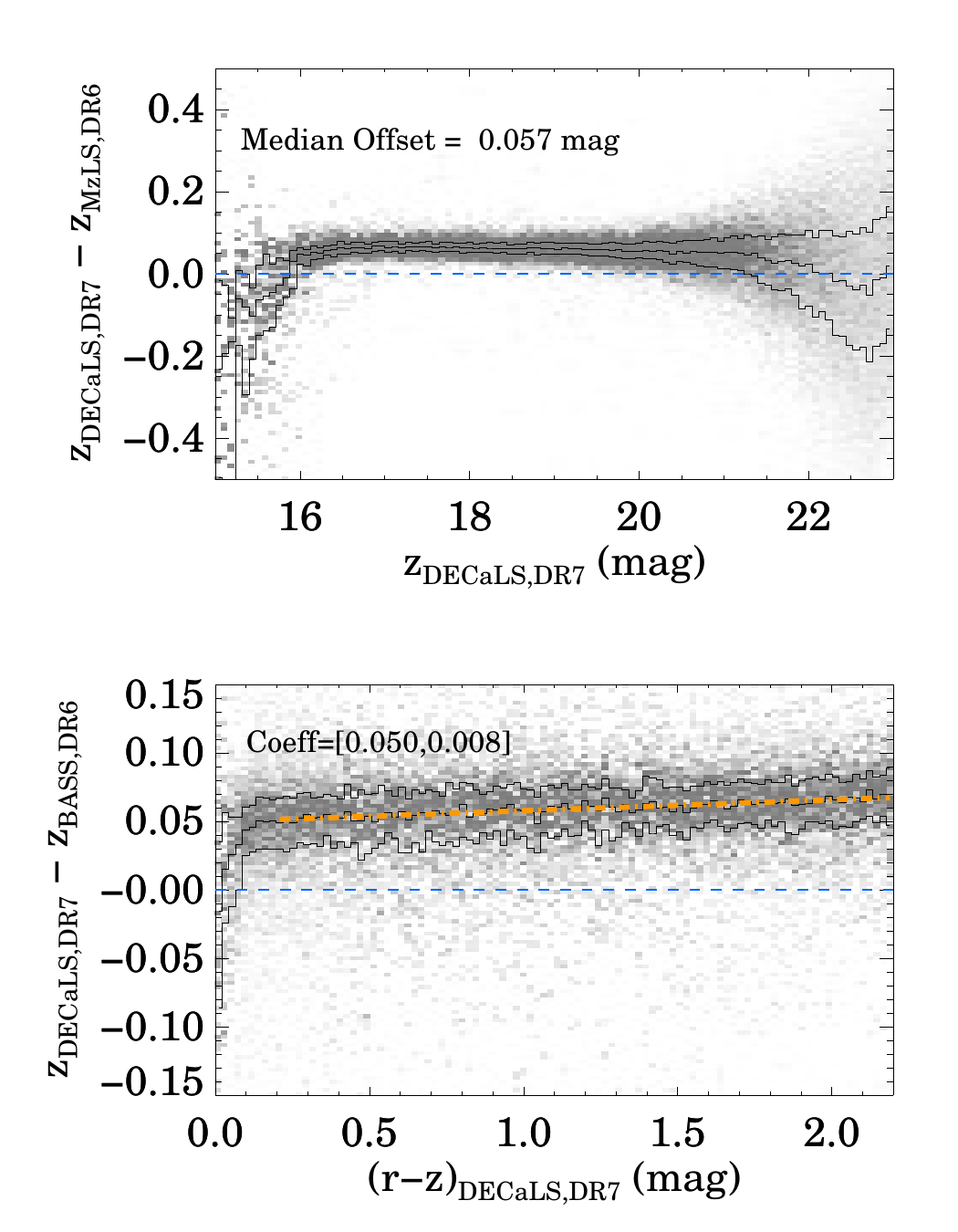}
\caption{Comparison of $grz$ photometry (top panels) and color transformations (bottom panels) for unresolved, unmasked, high signal-to-noise ratio sources in common between the Legacy Surveys DR7 (DECaLS) and DR6 (BASS and MzLS) releases. In each panel the zero level is shown by a blue dashed line; the solid black lines represent the 25\%, median and 75\% quantiles, and the orange dot-dashed curve shows the polynomial fit. While there are significant color terms between the different telescope+camera combinations, the median differences are $<$30\,mmag. The zero point offsets between the sub-surveys is due to a change in the approach to absolute calibration between DR6 and DR7; future releases will be self consistent.
\label{fig:dr5dr6color}}
\end{figure}

Figures~\ref{fig:skyzpt}, \ref{fig:skyastrom} and \ref{fig:zpsf} show, respectively, the spatial distributions of the photometric residuals in the $g$, $r$, and $z$ bands relative to the PS1 survey, the astrometric offsets relative to the Gaia DR1 catalog positions, and the best $z$-band image quality over the entire survey footprint. The three surveys provide nearly uniform coverage in photometry and astrometry, with rms scatter of $<$10~mmag and $<$5mas for well-measured stars. Nevertheless, there remain systematic offsets between the subsurveys; these are particularly obvious at the boundary between the DECaLS and MzLS/BASS surveys, at declination $\approx+34^\circ$ in the North Galactic Cap. The astrometric offset results from a change in the way in which the Gaia positions were included in DR7: while the astrometric calibration in both DR6 and DR7 are tied to Gaia DR1, Gaia stars with proper motions in Gaia DR2 were modeled by \tractor\ at the epoch of observation. Future Legacy Surveys data releases will be consistently tied to the latest version of the Gaia catalog, and self-consistently account for stars with well-measured proper motions.  

Despite efforts to keep the filter bandpasses as similar as possible for the Kitt Peak and Cerro Tololo surveys, there remain significant differences in the effective 
throughput as a function of wavelength between the different surveys, especially in the $g$ and $r$ photometric bands (see figure~\ref{fig:filters}). 
 A direct comparison of the photometry for stellar sources observed by both sets yields the following color transformations (see Figure~\ref{fig:dr5dr6color}):
\begin{eqnarray}
g_{\rm DR6,BASS} &=& g_{\rm DR7,DECaLS} -0.039 + 0.038(g-r)_{\rm DR7} + 0.195(g-r)_{\rm DR7}^2 - 0.251(g-r)_{\rm DR7}^3 \\ \nonumber
& & + 0.092(g-r)_{\rm DR7}^4 \\
r_{\rm DR6,BASS} &=& r_{\rm DR7,DECaLS} -0.004 + 0.035(r-z)_{\rm DR7} - 0.004(r-z)_{\rm DR7}^2 \\
z_{\rm DR6,MzLS} &=& z_{\rm DR7,DECaLS} -0.050 - 0.008(r-z)_{\rm DR7}
\end{eqnarray}
These color transformations are determined using unresolved objects (i.e., type ``PSF'' in both catalogs) with no masked pixels which are well detected (SNR$\ge$5) within the overlapping regions in both the DECaLS DR7 and BASS+MzLS DR6 catalogs (at ${\rm 149.9<RA<220.1}$, ${\rm 31.9<DEC<33.6}$). The polynomial least-squares fits were performed on samples restricted in magnitude and color: 
[15,22] and $0<(g-r)_{\rm DR7}<1.7$ in $g$-band; and
[15,21.5] and $0<(r-z)_{\rm DR7}<2.2$ in $r$-band and $z$-band.
The photometry in the catalogs has not been corrected for these color transformations.

The median photometric zero point offsets between DR6 and DR7 are:
\begin{eqnarray}
<g_{\rm DR6,BASS} - g_{\rm DR7,DECaLS}> &=& -0.028\,{\rm mag} \\
<r_{\rm DR6,BASS} - r_{\rm DR7,DECaLS}> &=& -0.024\,{\rm mag} \\
<z_{\rm DR6,MzLS} - z_{\rm DR7,DECaLS}> &=& 0.057\,{\rm mag}
\end{eqnarray}
These medians are computed for unresolved sources (i.e., type `PSF' in both catalogs) with magnitudes and colors in the range [16,22] and $0<(g-r)<1.7$  for $g$-band; and [16,22] and $0<(r-z)<2.2$ in the $r$- and $z$-bands. Part of the zero point offset is due to a change in the way in which the photometry was absolutely calibrated between DR6 and DR7. All data releases prior to DR7 fixed the absolute calibration such that the PS1 and Legacy Surveys magnitudes agreed for stars at a PS1 color of $(g-i)_{\rm PS1}=0$, whereas DR7 and all future releases place the Legacy Surveys' photometry on the AB magnitude scale, where small magnitude differences are expected at $(g-i)_{\rm PS1}=0$. This change resulted in offset differences of +0.009, $-$0.012, and +0.043 mag for the $g$, $r$, and $z$-bands, respectively. In addition, some portion of the median offsets computed above may be due to differences in the way the sky background is estimated in DR6 and DR7. We are currently working on the sky estimation algorithms and these results may change before our final data release.
\begin{figure}[!th]
\includegraphics[width=0.9\textwidth]{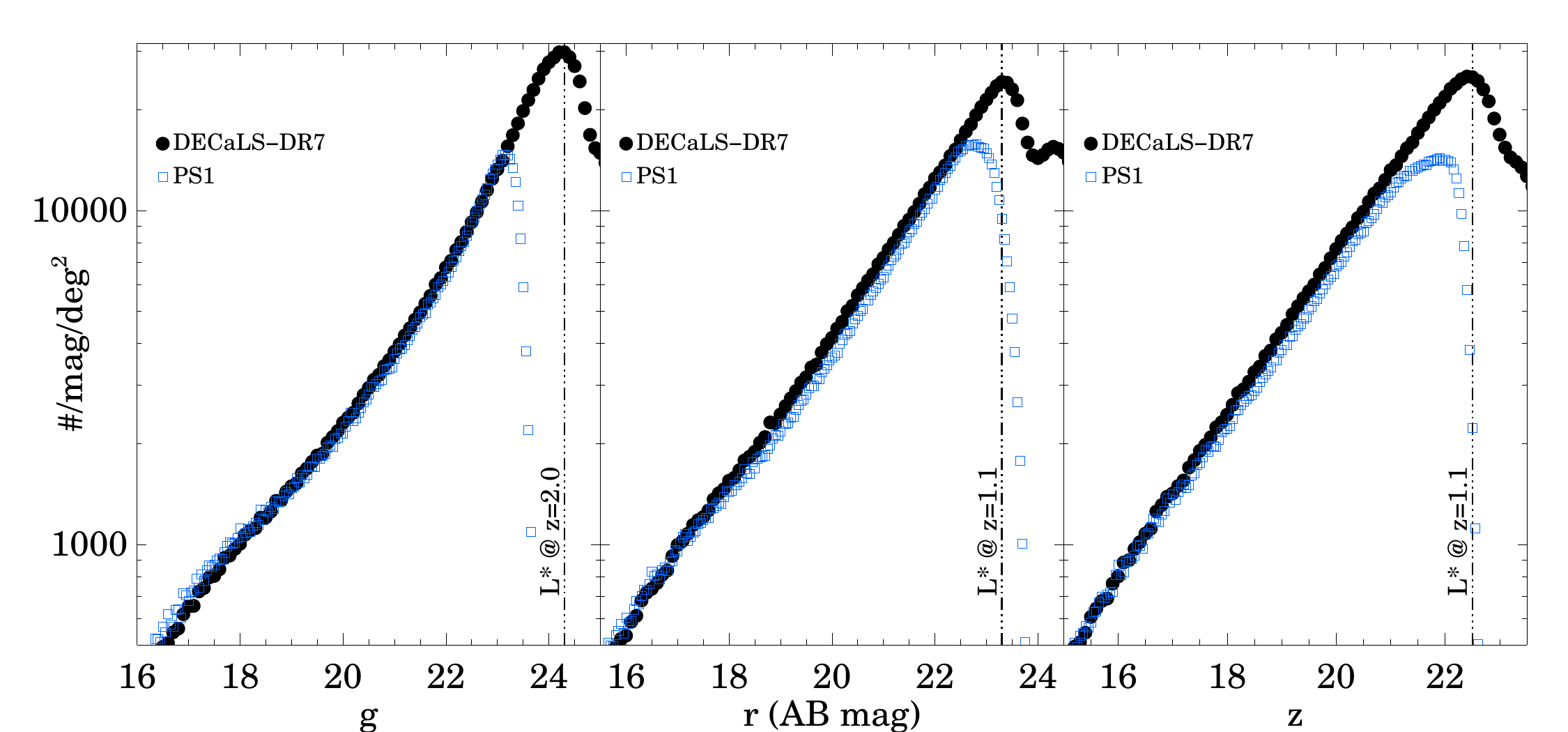}
\includegraphics[width=0.9\textwidth]{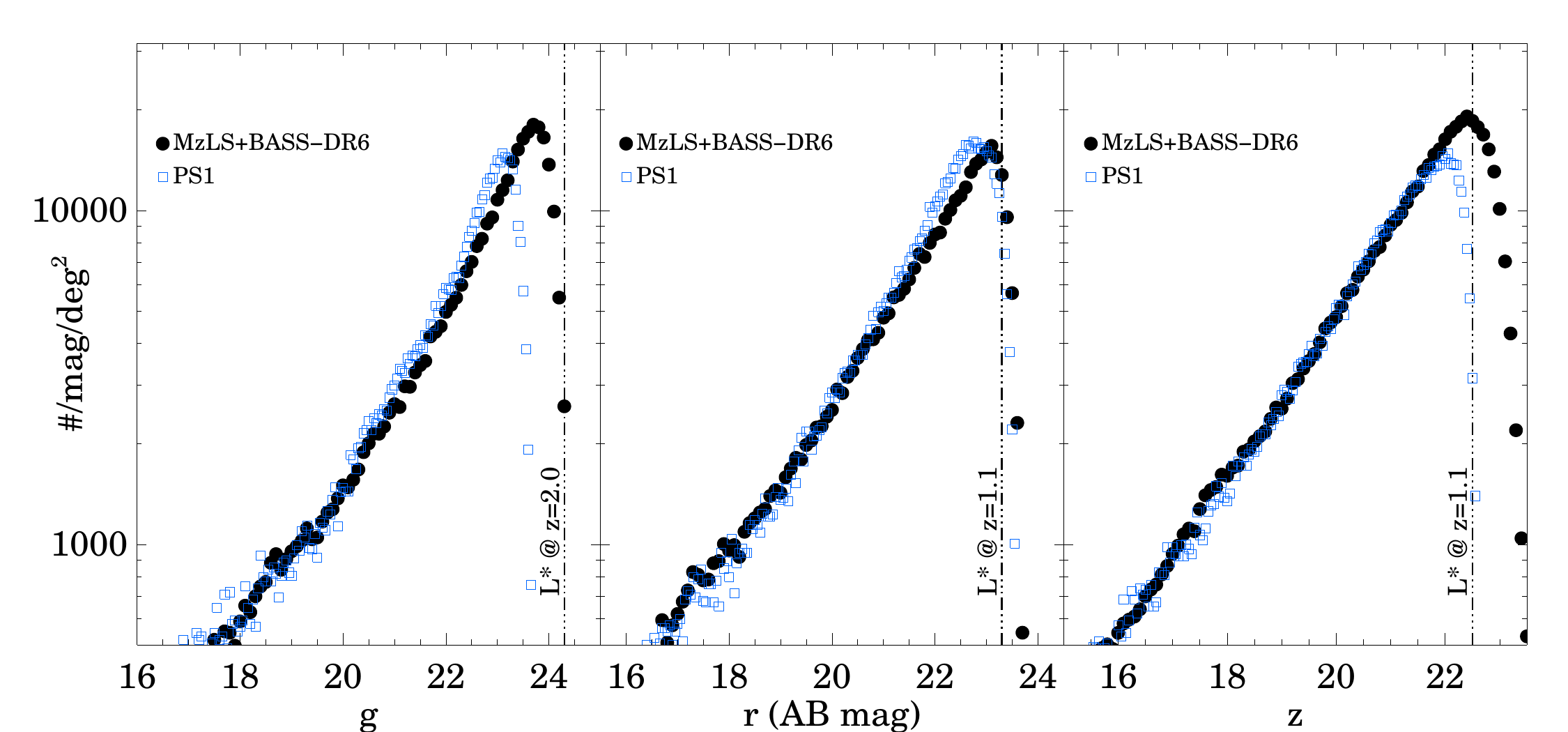}
\caption{Number counts of all detected sources measured from the Legacy Surveys DR6 and DR7 data releases compared with source counts from the PS1 survey. The DR7 number counts are measured in an $\approx20.4$\,deg$^2$ region centered on (RA,DEC)=(243.311$^\circ$,9.387$^\circ$). The DR6 number counts are measured in an $\approx 3.3$\,deg$^2$ region centered on (245.55$^\circ$,43.27$^\circ$).  All the data shown are for sources detected at signal-to-noise ratios $\ge5$, and the PS1 sources are required to be detected in at least two of the $grz$ bands in order to exclude spurious sources. 
\label{fig:numbercounts}}
\end{figure}

\begin{figure}[!th]
\includegraphics[width=1.0\textwidth]{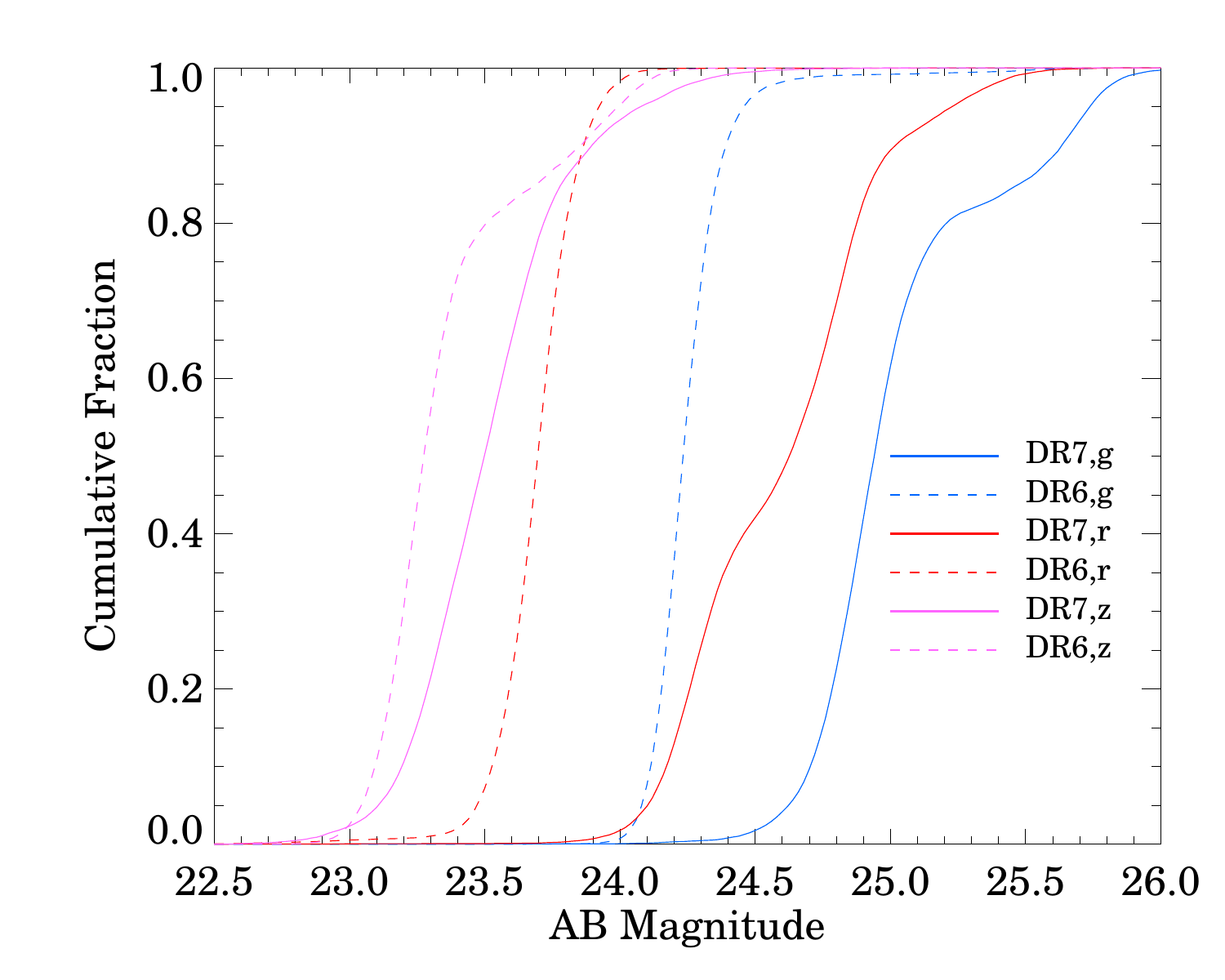}
\caption{The $grz$ 5$\sigma$ point-source depths in $\ge$\,3-pass data in the Legacy Surveys DR7 (data from DECaLS survey) and DR6 (data from BASS+MzLS surveys) releases.\label{fig:depths}}
\end{figure}

\begin{figure}[!th]
\includegraphics[width=1.0\textwidth]{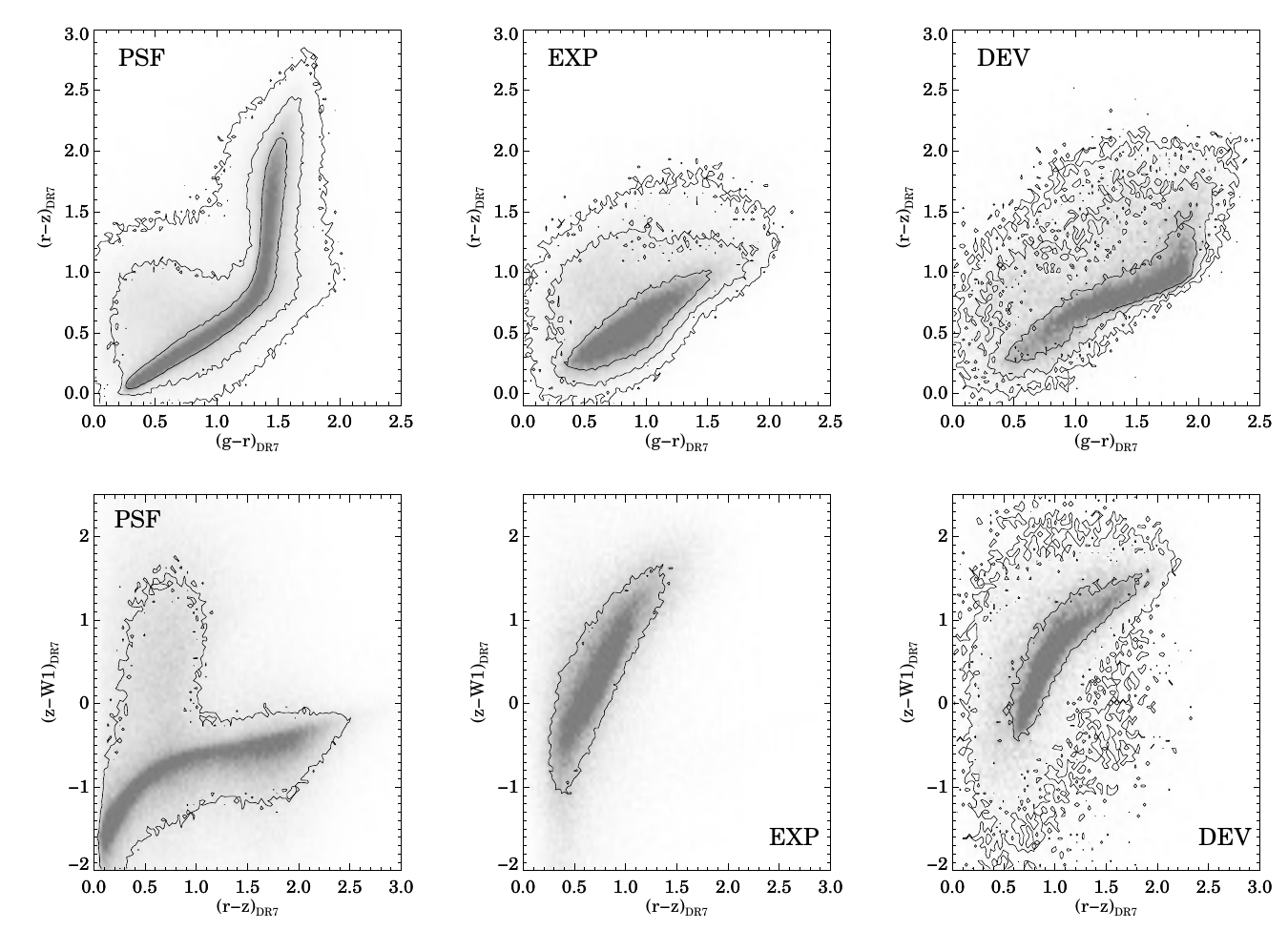}
\caption{The distribution, in the $g-r-z-W1$ color-color planes, of three of the morphological model types from DR7. The PSF type (left panels) corresponds to unresolved sources, which include stars, distant QSOs and some unresolved galaxies. The EXP and DEV types (middle and right panels respectively) are sources that are best fit by exponential disk or de Vaucouleurs $r^{1/4}$ light profiles. All colors are in units of AB magnitudes. \label{fig:colorcolor}}
\end{figure}

Figure~\ref{fig:numbercounts} shows the $grz$-band source number counts derived from DR6 and DR7 of the Legacy Surveys. Figure~\ref{fig:depths} shows the $grz$-band depths for the DR6 and DR7 data. The DR6 imaging (from BASS and MzLS) is shallower than the DR7 imaging (from DECaLS), but still satisfies the DESI target selection requirements.

Figure~\ref{fig:colorcolor} shows the distributions, in $grzW1$ color-color planes, of sources from the Legacy Surveys DR7 differentiated by their morphological type. Sources best fit by PSF models are dominated by stars (as exemplified by the clearly visible stellar locus), but also contain compact galaxies and QSOs. The sources best fit by spatially extended models (EXP: exponential disks; and DEV: de Vaucouleurs $r^{1/4}$ profiles) trace the galaxy locus with minimal contamination from stars (since the star-galaxy separation is limited by the ground-based seeing).

\begin{figure}[!ht]
\begin{centering}
\includegraphics[width=1.0\textwidth]{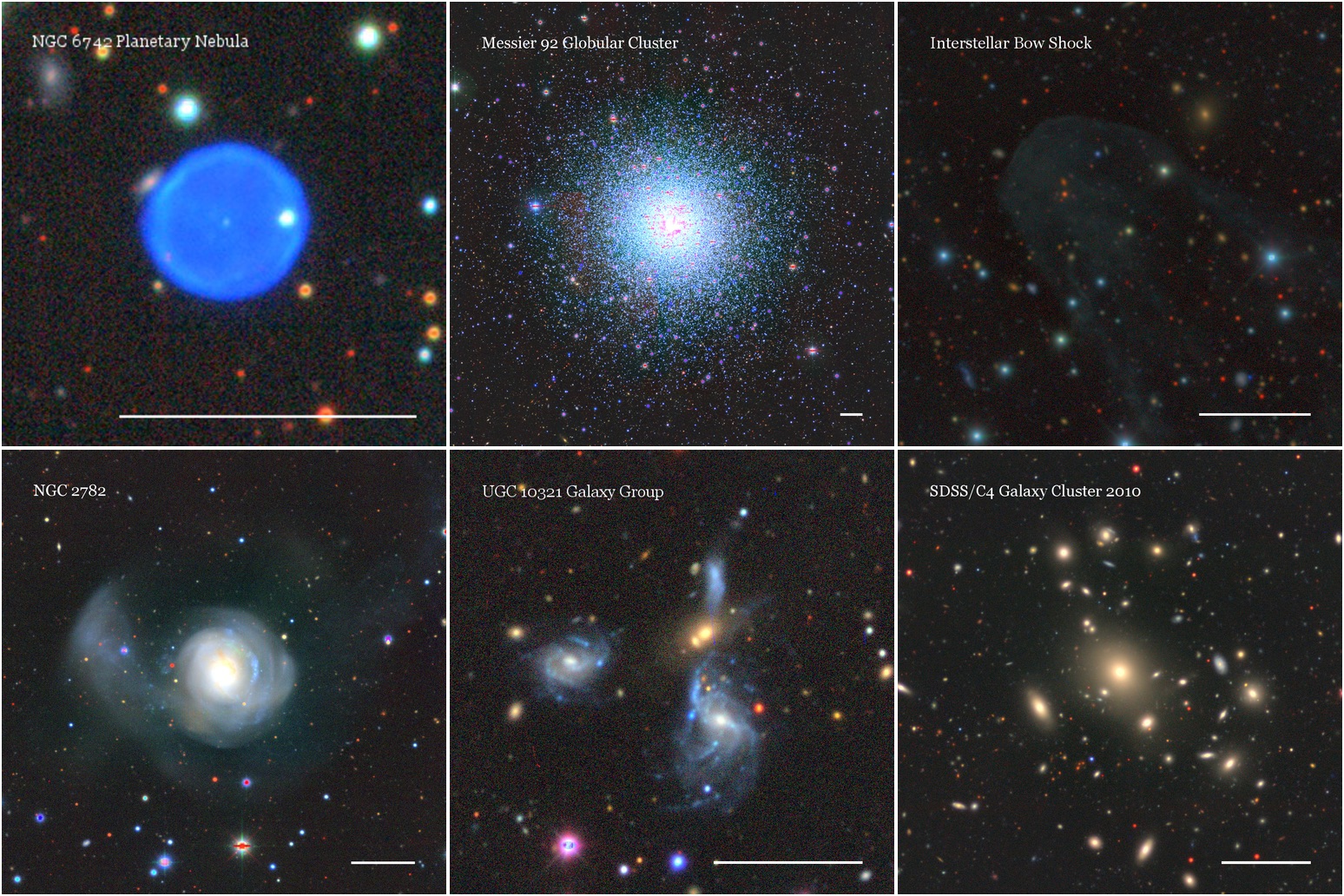}
\end{centering}
\caption{A gallery of image cutouts from the DR6 Legacy Surveys data illustrating the variety of astronomical objects covered by the surveys and highlighting the capability of the surveys to image low surface brightness features. The horizontal bar in the bottom right corner of each panel represents an angular size scale of 1\arcmin. In all images, north is up and east is to the left. More examples are shown in the DR6 and DR7 online \href{http://portal.nersc.gov/project/cosmo/data/legacysurvey/dr6/gallery/}{image galleries}.\label{fig:gallery}}
\end{figure}

\begin{figure}[!th]
\begin{centering}
\includegraphics[width=0.45\textwidth]{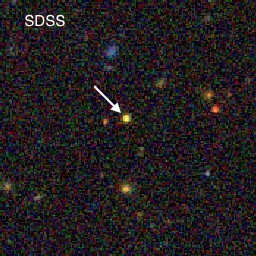}
\includegraphics[width=0.45\textwidth]{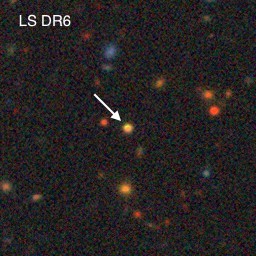}
\end{centering}
\caption{A new high proper motion star at (RA,DEC)=(223.7678$^\circ$,+33.7066$^\circ$) discovered serendipitously in DR6 of the Legacy Surveys. The left and right panels show $\approx67\times67$\arcsec\ cutouts of the SDSS and DR6 images, respectively, extracted from the Legacy Surveys' \href{http://legacysurvey.org/viewer}{image viewer}; north is up and east is to the left. The $g=21.2$\,AB\,mag M0V star (classified based on an \href{https://dr12.sdss.org/spectrumDetail?mjd=54942&fiber=500&plateid=3315}{SDSS spectrum}) 
is moving at $\approx$0.2\arcsec\ per year. The large time baseline ($\approx$14 years) between the SDSS and Legacy Surveys data will enable proper motion measurements for stars fainter than the Gaia DR1 catalog limits. \label{fig:HPMstar}}
\end{figure}

Figure~\ref{fig:gallery} shows cutouts of a wide variety of astronomical sources imaged by the Legacy Surveys, to demonstrate the depth and image quality of the surveys. The large time baseline between the Legacy Surveys and SDSS also enable the detection of faint transients and the measurement of proper motions for faint stars (Figure~\ref{fig:HPMstar} shows an example). The sensitivity of the Legacy Surveys data to low surface brightness structures has been exploited in the new version of Galaxy Zoo (\url{https://blog.galaxyzoo.org/tag/decals/}) and for finding faint tidal features associated with galaxies \citep[e.g.,][]{hood2018}. 
In addition, the depth of the Legacy Surveys data has resulted in their use for selecting emission-line galaxy \citep[DECaLS component;][]{raichoor2017} and QSO targets \citep[{\it WISE} component][]{Mye15} for the SDSS-IV/eBOSS spectroscopic survey.

All observations for the MzLS completed on 2018 February 12. Observing for BASS will continue through 2018 July. Observing for DECaLS is estimated to continue until 2019 March. The goal of the Legacy Surveys is to consolidate the data from all the subsidiary surveys into a final data release by 2019 June. 

\begin{table}[ht]
\centering
\caption{Data Releases for the Legacy Surveys}
\label{tab:drs}
\begin{tabular}{clcccc}
\hline
 DR & Release Date & Surveys & W1/W2  & Image/Catalog & Number of \\
    &              &        &     Depth (yr)          & Data Volume & Sources \\
  \hline
 1 & 2015 May & DECaLS & 1 & 15TB/238GB & 140 million\\ 
 2 & 2016 January & DECaLS & 2 & 33TB/301GB & 260 million\\  
 3 & 2016 September & DECaLS & 2 & 57TB/663GB & 478 million\\
 4 & 2017 June & BASS+MzLS & 3 & 30TB/256GB & 183 million\\
 5 & 2016 October & DECaLS & 3 & 18TB/969GB & 680 million\\
 6 & 2018 February & BASS+MzLS & 4 & 9TB/449GB & 310 million\\
 7 & 2018 July &  DECaLS & 4 & 21TB/1.5TB & 835 million\\ 
\hline
 8 & 2019 February & All & 5 & & \\
 9 & 2019 June & All & 5 & & \\
 \hline
\end{tabular}
\end{table}

\begin{table}[ht]
\begin{center}
\caption{Data Products from the Legacy Surveys}
\label{tab:dataprods}
\begin{tabular}{lcl}
\hline
{\bf Data Product}  & {\bf Type}  & {\bf Description}  \\
{\bf Name} & & \\
\hline
image  &  image & Coadded image per band$^1$ \\
invvar  &  image & Inverse-variance map per band \\
model  &  image & Coadded model image per band \\
chi2  &  image & Coadded chi-squared image ((image-model)$^2\times$invvar) per band \\
depth  &  image & PSF depth (as inverse-variance) per band \\
galdepth  &  image & Fiducial galaxy depth (as inverse-variance) per band \\
nexp & image & Number of exposures per band \\
tractor & catalog & Tractor catalog of measured sources, per brick \\
 \hline
\end{tabular}
\end{center}
$^1$ The released coadded images are inverse-variance-weighted coadditions of the individual reduced and calibrated images per brick, where the individual images have not been convolved to a common PSF. \\
\end{table}

%
%

\section{Summary \label{sec:summary}}

We have begun three wide-area optical imaging surveys, that will cover a total area of 14,000\,deg$^2$ in two contiguous portions and in three filters. These surveys are designed to provide targets for the DESI project, which will begin spectroscopic observations in 2019. The surveys include optical imaging data in three bands ($g$-,$r$-, and $z$-band) from three ground-based telescopes (CTIO Blanco, KPNO Mayall and Steward Bok 90") and will reach approximate $5\sigma$ depths of $g$=24.0, $r$=23.4 and $z$=22.5\,AB\,mag for faint galaxies. In addition, the data releases will contain mid-infrared photometry based on new, multi-epoch stacks of imaging data from the {\it WISE} satellite created by the unWISE project.  The three surveys are the Dark Energy Camera Legacy Survey (DECaLS), the Mayall $z$-band Legacy Survey (MzLS) and the Beijing-Arizona Sky Survey (BASS); these surveys are jointly referred to as the Legacy Surveys. 

We have implemented an automated observing strategy for all surveys, where observing conditions are monitored in near-real time and exposure times are modified and target field selection is optimized on-the-fly to ensure that each individual observation reaches the required depth. This strategy ensures that the ground-based surveys produce datasets that are as uniform as possible.  The data are pipeline processed using the NOAO Community Pipelines, and catalogs are then generated from the imaging data using an inference-based forward modeling approach known as \tractor, implemented at NERSC.  

All the imaging data, source catalogs, and code from the  Legacy Surveys are publicly available: the raw and pipeline-processed optical imaging data are made public as soon as they are available in the NOAO Archive and catalogs from the project are released approximately twice  each year. There have been 7 data releases thus far, which are described in detail in other papers. The imaging data are accessible from the 
Legacy Surveys website\footnote{\url{http://legacysurvey.org}, served by NERSC} and the NOAO Science Archive\footnote{\url{http://archive.noao.edu}}. An image viewer (the {\it Imagine} sky viewer, at \url{http://legacysurvey.org/viewer}) is provided at the Legacy Surveys website that allows the user to examine the data,  \tractor\ models, and other supplementary data. A resource at the NOAO DataLab provides access to the catalogs through SQL queries or a Jupyter notebook interface\footnote{see \url{https://datalab.noao.edu/decals/ls.php} for details}.

The Legacy Surveys are nearing completion. We anticipate completing the surveys before the start of DESI spectroscopic observations in late 2019. 

\acknowledgements

This paper presents observations obtained at Cerro Tololo Inter-American Observatory, National Optical Astronomy Observatory (NOAO Prop. ID: 2014B-0404; co-PIs: D.\ J.\ Schlegel and A.\ Dey), which is operated by the Association of Universities for Research in Astronomy (AURA) under a cooperative agreement with the National Science Foundation. This paper also includes DECam observations obtained as part of other projects, namely 
the Dark Energy Survey (DES, NOAO Prop. ID: 2012B-0001); 
and 2012B-0003, 2012B-0416, 2012B-0506, 2012B-0569, 2012B-0617, 2012B-0621, 2012B-0624, 2012B-0625, 2012B-3003, 2012B-3011, 2012B-3012, 2012B-3016, 2012B-9993, 2012B-9999, 2013A-0327, 2013A-0360, 2013A-0386, 2013A-0400, 2013A-0455, 2013A-0529, 2013A-0609, 2013A-0610, 2013A-0611, 2013A-0613, 2013A-0614, 2013A-0618, 2013A-0704, 2013A-0716, 2013A-0717, 2013A-0719, 2013A-0723, 2013A-0724, 2013A-0737, 2013A-0739, 2013A-0741, 2013A-9999, 2013B-0325, 2013B-0438, 2013B-0440, 2013B-0453, 2013B-0502, 2013B-0531, 2013B-0612, 2013B-0613, 2013B-0615, 2013B-0616, 2013B-0617, 2014A-0073, 2014A-0191, 2014A-0239, 2014A-0255, 2014A-0256, 2014A-0270, 2014A-0306, 2014A-0313, 2014A-0321, 2014A-0327, 2014A-0339, 2014A-0348, 2014A-0386, 2014A-0390, 2014A-0412, 2014A-0415, 2014A-0429, 2014A-0496, 2014A-0608, 2014A-0610, 2014A-0611, 2014A-0613, 2014A-0620, 2014A-0621, 2014A-0622, 2014A-0623, 2014A-0624, 2014A-0632, 2014A-0640, 2014B-0146, 2014B-0244, 2014B-0608, 2014B-0610, 2014B-0614, 2015B-0187. 

DECaLS used data obtained with the Dark Energy Camera (DECam), which was constructed by the Dark Energy Survey (DES) collaboration. Funding for the DES Projects has been provided by the U.S. Department of Energy, the U.S. National Science Foundation, the Ministry of Science and Education of Spain, the Science and Technology Facilities Council of the United Kingdom, the Higher Education Funding Council for England, the National Center for Supercomputing Applications at the University of Illinois at Urbana-Champaign, the Kavli Institute of Cosmological Physics at the University of Chicago, Center for Cosmology and Astro-Particle Physics at the Ohio State University, the Mitchell Institute for Fundamental Physics and Astronomy at Texas A\&M University, Financiadora de Estudos e Projetos, Funda{\c c}{\~a}o Carlos Chagas Filho de Amparo, Financiadora de Estudos e Projetos, Funda{\c c}{\~a}o Carlos Chagas Filho de Amparo {\`a} Pesquisa do Estado do Rio de Janeiro, 
Conselho Nacional de Desenvolvimento Cient{\'i}fico e Tecnol{\'o}gico and the Minist{\'e}rio da Ci{\^e}ncia, Tecnologia e Inovac{\~a}o, the Deutsche Forschungsgemeinschaft and the Collaborating Institutions in the Dark Energy Survey. 
The Collaborating Institutions are Argonne National Laboratory, the University of California at Santa Cruz, the University of Cambridge, 
Centro de Investigaciones En{\'e}rgeticas, Medioambientales y Tecnol{\'o}gicas-Madrid, the University of Chicago, University College London, the DES-Brazil Consortium, the University of Edinburgh, 
the Eidgen{\"o}ssische Technische Hoch\-schule (ETH) Z{\"u}rich, 
Fermi National Accelerator Laboratory, the University of Illinois at Urbana-Champaign, 
the Institut de Ci{\`e}ncies de l'Espai (IEEC/CSIC), 
the Institut de F{\'i}sica d'Altes Energies, 
Lawrence Berkeley National Laboratory, 
the Ludwig-Maximilians Universit{\"a}t M{\"u}nchen and the associated Excellence Cluster Universe, 
the University of Michigan, the National Optical Astronomy Observatory, the University of Nottingham, the Ohio State University, the University of Pennsylvania, the University of Portsmouth, SLAC National Accelerator Laboratory, Stanford University, the University of Sussex, and Texas A\&M University. 

The Mayall $z$-band Legacy Survey (MzLS; NOAO Prop. ID \# 2016A-0453; PI: A.\ Dey) uses observations made with the Mosaic-3 camera at the Mayall 4m telescope at Kitt Peak National Observatory, National Optical Astronomy Observatory, which is operated by the Association of Universities for Research in Astronomy (AURA) under cooperative agreement with the National Science Foundation. The authors are honored to be permitted to conduct astronomical research on Iolkam Du'ag (Kitt Peak), a mountain with particular significance to the Tohono O'odham. 

The Beijing-Arizona Sky Survey (BASS; NOAO Proposal ID \# 2015A-0801; PIs: Zhou Xu and Xiaohui Fan) is a key project of the Telescope Access Program (TAP), which has been funded by the National Astronomical Observatories of China, the Chinese Academy of Sciences (the Strategic Priority Research Program ``The Emergence of Cosmological Structures" Grant \# XDB09000000), and the Special Fund for Astronomy from the Ministry of Finance. The BASS is also supported by the External Cooperation Program of Chinese Academy of Sciences (Grant \# 114A11KYSB20160057), and Chinese National Natural Science Foundation (Grant \# 11433005). The Bok Telescope is located on Kitt Peak and operated by Steward Observatory, University of Arizona.

The Legacy Surveys imaging of the DESI footprint is supported by the Director, Office of Science, Office of High Energy Physics of the U.S.\ Department of Energy under Contract No. DE-AC02-05CH11231, by the National Energy Research Scientific Computing Center, a DOE Office of Science User Facility under the same contract; and by the U.S.\ National Science Foundation, Division of Astronomical Sciences under Contract No.\ AST-0950945 to NOAO. Travel and other support for the DECaLS and MzLS projects is provided by the National Optical Astronomy Observatory, the Lawrence Berkeley National Laboratory, and the DESI Project.

This publication makes use of data from the Pan-STARRS1 Surveys (PS1) and the PS1 public science archive, which have been made possible through contributions by the Institute for Astronomy, the University of Hawaii, the Pan-STARRS Project Office, the Max-Planck Society and its participating institutes, the Max Planck Institute for Astronomy, Heidelberg and the Max Planck Institute for Extraterrestrial Physics, Garching, The Johns Hopkins University, Durham University, the University of Edinburgh, the Queen's University Belfast, the Harvard-Smithsonian Center for Astrophysics, the Las Cumbres Observatory Global Telescope Network Incorporated, the National Central University of Taiwan, the Space Telescope Science Institute, the National Aeronautics and Space Administration under Grant No. NNX08AR22G issued through the Planetary Science Division of the NASA Science Mission Directorate, the National Science Foundation Grant No. AST-1238877, the University of Maryland, Eotvos Lorand University (ELTE), the Los Alamos National Laboratory, and the Gordon and Betty Moore Foundation.

This work has made use of data from the European Space Agency (ESA)
mission {\it Gaia} (\url{https://www.cosmos.esa.int/gaia}), processed by
the {\it Gaia} Data Processing and Analysis Consortium (DPAC,
\url{https://www.cosmos.esa.int/web/gaia/dpac/consortium}). Funding
for the DPAC has been provided by national institutions, in particular
the institutions participating in the {\it Gaia} Multilateral Agreement.

This publication makes use of data products from the Wide-field Infrared Survey Explorer (and its successor,  the Near-Earth Object Wide-field Infrared Survey Explorer),  which is a joint project of the University of California, Los Angeles, and the Jet Propulsion Laboratory/California Institute of Technology, funded by the National Aeronautics and Space Administration.

This research used resources of the National Energy Research Scientific Computing Center, a DOE Office of Science User Facility supported by the Office of Science of the U.S. Department of Energy under Contract No. DE-AC02-05CH11231.

The research leading to these results has received funding from the
European Research Council under the European Union's Seventh Framework
Programme (FP/2007-2013) / ERC Grant Agreement n. 320964 (WDTracer).

J. Moustakas gratefully
acknowledges support from the National Science Foundation grant
AST-1616414. A. Dey thanks the Radcliffe Institute for Advanced Study for their generous support during the year these surveys were initiated. A. Dey, D. J. Schlegel, and D. Lang thank the Aspen Center for Physics, which is supported by National Science Foundation grant PHY-1066293, for their hospitality and support during summer 2015 during which part of this work was conducted. 

\appendix

\section{DESI Imaging Requirements \label{sec:desirequirements}}

Imaging data from the Legacy Surveys will be used to select targets for DESI. In this Appendix, we describe the basic requirements that
the DESI project's main cosmological survey imposes on imaging.
There are five target classes for the DESI dark energy experiment: (1) Bright Galaxy Sample (BGS); 
(2) Luminous Red Galaxies (LRGs); (3) Emission Line Galaxies (ELGs); (4) quasars (QSOs) at $z<2.1$;  and (5) quasars at $z>2.1$ for the measurement of the Lyman-alpha forest.
The imaging surveys will also be used to select secondary targets such as other interesting object classes (rare objects, Milky Way stars, etc.), standard stars for
spectrophotometric calibration, and locations for blank sky fibers.

In 2013, the DESI Project concluded that a three-band $g/r/z$ optical imaging program complemented by {\it WISE} W1 and W2 photometry would be sufficient  to select all target classes required for  the DESI cosmology program. 
One optical band is sufficient to select the BGS, which is defined as a simple
magnitude-limited sample. Two optical bands are necessary to select LRGs and quasars, as the {\it WISE} W1 infrared band
provides adequate color information to clearly separate these targets from
stars and other galaxies.
Three optical bands are necessary to efficiently select the ELGs,
as these galaxies are not well-detected at the depth of the {\it WISE} data.
Other imaging data (e.g., $u$-band) may be used to further refine the selection of the high-redshift QSOs,
which is the one target class that is not required (by DESI) to have a uniform
selection if used for Lyman-alpha forest maps.

The requirements that DESI targeting imposes on the optical imaging are:
\begin{enumerate}
\item {\bf Imaging will be in three optical bands to a depth of at least
$g=24.0$, $r=23.4$ and $z=22.5$.}
The depths are defined as the optimal-extraction (forced-photometry)
depths for a galaxy near the depth limits of DESI, where that galaxy
is defined to be an exponential profile with a half-light radius
of $r_{\rm half} = 0.45$\,arcsec.
For such a profile, the effective number of pixels is well-approximated by
$N_{\rm eff} = \left[ (4\pi\sigma^2)^{(1/p)} + (8.91 r_{\rm half}^2)^{(1/p)} \right]^p$,
where $\sigma$ is the standard deviation for a Gaussian fit to the seeing,
$r_{\rm half}$ is the half-light radius for an exponential-profile galaxy,
and $p=1.15$.
In addition to the optical imaging, 4 years of {\it WISE} data in the W1 and W2 bands are assumed.
\item {\bf Imaging will cover at least 14,000~deg$^2$ of the DESI footprint.}  This footprint is described in $\S$\,\ref{sec:footprint}.
\item {\bf The fill factor will be at least 90\%}.
The areas with coverage to full depth in all 3 bands should exceed 90\% of
the footprint. The science loss is approximately proportional to this
fractional loss of area.
\item {\bf $z$-band image quality will be smaller than 1.5\,arcsec FWHM.}
Many of the DESI galaxy targets appear small (less than 1\,arcsec)
and are near the limit at which they can be resolved as extended sources
rather than point sources. Because of this, a morphological
separation between extended sources and point sources will not be
required for DESI galaxy targets.
The primary driver for reasonably good image quality is to minimize
blending between targets and other sources on the sky, especially
in regions of high stellar density, such as near the Galactic plane. 
Reasonable image quality in at least one band will allow the identification
of otherwise blended objects and the optimal extraction of their photometry
in all bands. All of the DESI targets will be detected in $z$-band, which is 
expected to have the best image quality, therefore an
image quality requirement is specified in that one band.
\item {\bf No regions larger than 3\,deg in diameter will be covered by only nonphotometric
observations.}
Photometric observations are defined as those obtained during nights
demonstrated to have photometricity errors of 1\% RMS in $g$-band and $r$-band,
or 2\% RMS in $z$-band. Small regions, especially in CCD gaps on the imager
focal planes, may be filled in with nonphotometric data and calibrated
to neighboring, photometric data.
\item {\bf The random errors in astrometry will be less than 95\,mas RMS,
and the systematic errors will be less than 30\,mas.}
Astrometric errors impact the effective throughput of the DESI instrument
for the spectroscopic survey.
The systematics errors in the astrometry will be well-controlled by
tying individual CCD images to Gaia \citep{gaia}.
The astrometric positions can take advantage of the signal in all filters.
\end{enumerate}

An inference modeling approach will be used for measuring the photometry
in all bands from the optical and {\it WISE} imaging data.
The DESI requirements for the imaging {\em reductions} are:
\begin{enumerate}
\item {\bf The model photometry will use the same model in all bands.}
The model photometry uses the same model in all bands to minimize
systematic errors in the flux ratios (i.e., colors) of galaxies,
which is critical for all target selection algorithms.
\item {\bf Systematic errors due to PSF mis-estimation will be
controlled to better than 1\%}.
\item {\bf Systematic errors due to galaxy model mis-estimation will be
controlled to better than 1\%.}
A simple $\chi^2$ fit would suffer from biases in the galaxy photometry
due to the models not matching the actual morphology of individual
galaxies, and this would be S/N-dependent.
\item {\bf Depth maps will be computed in each band at all locations.}
The effective depth will be computed as a map for each filter.
This is a function of the sky brightness, image quality and sky
transparency of all of the contributing images. In detail on small
scales, it would include the features seen from bad columns on the
CCD and increased noise near other sources on the sky.
\item {\bf The ability will be provided to Monte Carlo simulated objects
through the imaging pipeline.} This requirement implies access to all
of the calibrated image frames, the versioned code used to construct
the final targeting catalogs, and the ability to run this code.
This Monte Carlo ability will be used to map the effects of bright
stars and other source contamination problems, which can both produce
spurious targets and remove actual targets by blending them with other sources. 
\end{enumerate}


\section{Transformations from SDSS to Legacy Surveys Photometry \label{sec:sdssvsls}}

We compared the SDSS and Legacy Surveys photometry in two regions using the ``sweeps'' files which contain matched SDSS and Legacy Survey sources and are included as part of the 
DR6 and DR7 Legacy Survey releases. 
For the SDSS-DR6 comparison, we used the region $230^\circ\le{\rm RA}\le240^\circ$ and $2.0^\circ\le{\rm DEC}\le 2.5^\circ$; for the SDSS-DR7 comparison, we used the region $160^\circ\le{\rm RA}\le170^\circ$ and $2.0^\circ\le{\rm DEC}\le 2.5^\circ$.
The photometric transformations were then computed by fitting polynomials to the SDSS-Legacy Survey magnitude differences in the $g$, $r$ and $z$ bands as a function of the SDSS $(g-i)$ color. The fits were done in the color range $-0.25\le(g-i)_{\rm SDSS}\le3.5$ and restricted to bright stars (i.e., $i\le19$). The resulting transformations are:
\begin{eqnarray}
g_{\rm DR6,BASS} &\approx& g_{\rm SDSS} -0.0125-0.0535(g-i)_{\rm SDSS}+0.0162(g-i)_{\rm SDSS}^2-0.0047(g-i)_{\rm SDSS}^3 \\
r_{\rm DR6,BASS} &\approx& r_{\rm SDSS} -0.0215-0.0683(g-i)_{\rm SDSS}+0.0265(g-i)_{\rm SDSS}^2-0.0084(g-i)_{\rm SDSS}^3\\
z_{\rm DR6,MzLS} &\approx& z_{\rm SDSS} -0.0293-0.0387(g-i)_{\rm SDSS}+0.0123(g-i)_{\rm SDSS}^2-0.0034(g-i)_{\rm SDSS}^3\\
g_{\rm DR7,DECaLS} &\approx& g_{\rm SDSS}+0.0244-0.1183(g-i)_{\rm SDSS}+0.0322(g-i)_{\rm SDSS}^2-0.0066(g-i)_{\rm SDSS}^3\\
r_{\rm DR7,DECaLS} &\approx& r_{\rm SDSS}-0.0005-0.0868(g-i)_{\rm SDSS}+0.0287(g-i)_{\rm SDSS}^2-0.0092(g-i)_{\rm SDSS}^3\\
z_{\rm DR7,DECaLS} &\approx& z_{\rm SDSS}+0.0228-0.0229(g-i)_{\rm SDSS}+0.0049(g-i)_{\rm SDSS}^2-0.0019(g-i)_{\rm SDSS}^3\\
\end{eqnarray}
The typical root-mean-square photometric scatter around these transformation relations is $\sigma\approx30$~mmag. These transformations are not valid for stars bluer or redder than the color range of the fit. 



\section{Comparing Subaru Hyper-Suprime Camera and Legacy Surveys Data \label{sec:hscvsls}}

We have compared data from DR1 of the Subaru Hyper-SuprimeCam (HSC) surveys \citep{HSC_SSP_DR1}\footnote{See \url{http://hsc.mtk.nao.ac.jp/ssp/survey/}} with data from the Legacy Surveys in two regions: 
(1) $244.5^\circ\le{\rm RA}\le246.5^\circ$, $43^\circ\le{\rm DEC}\le44^\circ$; and 
(2) $31.5^\circ\le{\rm RA}\le 36^\circ$, $-6^\circ\le{\rm DEC}\le-4.5^\circ$. 
The first region corresponds to the ``HECTOMAP'' region in the HSC survey, and is covered by the Legacy Surveys DR6 data. The second region corresponds to the XMM-LSS region in the HSC survey, and is covered by the Legacy Surveys DR7 data. Selecting Legacy Surveys sources of type 'PSF' detected with a signal-to-noise ratio $\ge$5, we compared the HSC photometry to the DR6 and DR7 data and derived the following transformations:
\begin{eqnarray}
g_{\rm DR6,BASS} &\approx& g_{\rm HSC}-0.003+0.029(g-r)_{\rm HSC}\\
r_{\rm DR6,BASS} &\approx& r_{\rm HSC}+0.003-0.130(r-z)_{\rm HSC}+0.053(r-z)_{\rm HSC}^2-0.013(r-z)_{\rm HSC}^3\\
z_{\rm DR6,MzLS} &\approx& z_{\rm HSC}-0.011-0.076(r-z)_{\rm HSC}+0.003(r-z)_{\rm HSC}^2\\
g_{\rm DR7,DECaLS} &\approx& g_{\rm HSC}+0.003-0.014(g-r)_{\rm HSC}\\
r_{\rm DR7,DECaLS} &\approx& r_{\rm HSC}-0.011-0.154(r-z)_{\rm HSC}+0.055(r-z)_{\rm HSC}^2-0.013(r-z)_{\rm HSC}^3\\
z_{\rm DR7,DECaLS} &\approx& z_{\rm HSC}-0.024-0.098(r-z)_{\rm HSC}+0.004(r-z)_{\rm HSC}^2\\
\end{eqnarray}
The 50\% completeness limits (measured relative to HSC) in [$g$,$r$,$z$] are [25.26,24.84,24.29]\,AB\,mag in DR7 and 
[24.535, 24.235, 23.336]\,AB\,mag in DR6, respectively. Of the 95,241 DR6 sources in region (1),  3,506 (3.7\%)  have no matches in the HSC catalog. 713 of these are bright sources ($g\le17,r\le18,z\le17$) and are likely saturated in the HSC data; 1,431 are fainter than the 80\% completeness limits. We examined 100 DR6 sources with magnitudes $20<g<22$ that are missing in the HSC catalogs and found that 83 are fake sources (33 due to particle events; 22 due to increased noise in, e.g., regions near CCD edges; 28 are fake sources created by \tractor\ in the extended halos of bright (typically saturated) stars or galaxies; and  the remaining are real sources that have been split into multiple sources due to their complexity, or mis-centered (due to low signal-to-noise data) in the DR6 catalog, or are low surface brightness sources not present in the HSC catalog. Despite these issues, the DR6 catalog reliability is higher than PS1 at faint magnitudes. 

\bibliographystyle{apj}
\bibliography{bibliography}

\facility{KPNO:Mayall (Mosaic-3)}
\facility{Steward:Bok (90Prime)}
\facility{CTIO:Blanco (DECam)}
\facility{WISE}
\facility{Gaia}

\end{document}